\documentclass[pre,twocolumn,aps,showpacs]{revtex4}
\newcommand{\bec}[1]{\mbox{\boldmath $ #1$}}
\usepackage{graphicx}
\begin{document}
\title{Effect of rotation on a developed turbulent stratified convection:
the hydrodynamic helicity, the alpha-effect and the effective drift velocity}
\author{Nathan Kleeorin}
\email{nat@menix.bgu.ac.il}
\author{Igor Rogachevskii}
\email{gary@menix.bgu.ac.il}
\homepage{http://www.bgu.ac.il/~gary}
\affiliation{Department of Mechanical Engineering, The Ben-Gurion
University of the Negev, \\
POB 653, Beer-Sheva 84105, Israel}
\date{\today}
\begin{abstract}
An effect of rotation on a developed turbulent stratified
convection is studied. Dependencies of the hydrodynamic helicity,
the alpha-tensor and the effective drift velocity of the mean
magnetic field on the rate of rotation and an anisotropy of
turbulent convection are found. It is shown that in an anisotropic
turbulent convection the alpha-effect can change its sign
depending on the rate of rotation. The evolution of the
alpha-effect is much more complicated than that of the
hydrodynamic helicity in an anisotropic turbulent convection of a
rotating fluid. Different properties of the effective drift
velocity of the mean magnetic field in a rotating  turbulent
convection are found: (i) a poloidal effective drift velocity can
be diamagnetic or paramagnetic depending on the rate of rotation;
(ii) there is a difference in the effective drift velocities for
the toroidal and poloidal magnetic fields; (iii) a toroidal
effective drift velocity can play a role of an additional
differential rotation. The above effects and an effect of a
nonzero divergence of the effective drift velocity of the toroidal
magnetic field on a magnetic dynamo in a developed turbulent
stratified convection of a rotating fluid are studied.
Astrophysical applications of the obtained results are discussed.
\end{abstract}


\maketitle

\section{Introduction}

Turbulent transport of particles and magnetic fields was
intensively studied for the Navier-Stokes turbulence (see, {\em
e.g.,} \cite{MY75,ZMR88,ZRS90,F95}). However, there are a number
of applications with other kinds of turbulence, {\em e.g.,}
turbulent convection. For instance, in the Sun and stars there is
a developed turbulent convection that is strongly influenced by a
fluid rotation.

The mean-field theory of magnetic field was in general developed
for the Navier-Stokes turbulence without taking into account
turbulent convection (see, {\em e.g.,}
\cite{M78,P79,KR80,ZRS83,RSS88,S89,RS92,R94}). In particular, the
dependencies of the the $ \alpha $-effect, the effective drift
velocity and the turbulent magnetic diffusion on the rate of
rotation were found only for the Navier-Stokes turbulence (see,
{\em e.g.,} \cite{K91,RK93,KPR94,RKR00}) in spite of that in many
astrophysical applications there are turbulent convection regions.
A turbulent convection in different situations has been studied
mainly by numerical simulations (see, {\em e.g.,}
\cite{ZB89,BN90,TB97,BGT98,B2000,OSB01,OSB02}).

In this paper we study an influence of rotation on a developed
turbulent stratified convection. This allows us to find the
dependencies of the hydrodynamic helicity, the alpha-tensor and the
effective drift velocity of the mean magnetic field on the rate of
rotation.

This study has a number of applications in astrophysics. In
particular, the evolution of the mean magnetic field in the
kinematic approximation (without taking into account a two-way
coupling of mean magnetic field and turbulent fluid flow) can be
described in terms of propagating waves with the growing
amplitude, i.e., the magnitude of the mean magnetic field $ B $ is
given by
\begin{eqnarray}
B \propto B_{0} \exp(\gamma_{_{B}} t) \cos(\omega_{_{B}} t
+ {\bf k} \cdot {\bf r})
\;,
\label{P1}
\end{eqnarray}
where $ B_{0} $ is a seed magnetic field, $ \gamma_{_{B}} $ is the
growth rate of the mean magnetic field, $ \omega_{_{B}} $ and $
{\bf k} $ are the frequency and the wave vector of a dynamo wave.
In the Sun, {\em e.g.,} according to the magnetic field
observations these dynamo waves with the $ \sim 22 $ years period
propagate to the equator (see, {\em e.g.,}
\cite{M78,P79,KR80,ZRS83,S89}). The magnetic field is generated in
the turbulent convective zone inside the Sun. The growth of the
mean magnetic field is a combined effect of a nonuniform fluid
rotation (the differential rotation, $ \bec{\nabla} (\delta
\Omega)) $ and helical turbulent motions (the $ \alpha$-effect).
The direction of propagation of the dynamo waves is determined by
a sign of the parameter $ \alpha [\partial (\delta \Omega) /
\partial r] ,$ where $ r, \theta , \varphi $ are the spherical
coordinates, and $ {\bf \Omega} $ is the angular velocity. When
the parameter $ \alpha \, [\partial (\delta \Omega) /
\partial r] $ is negative the dynamo waves propagates to
the equator. The helioseismology shows that in the solar
convective zone $ \partial (\delta \Omega) / \partial r > 0 $ and
the existing theories yield $ \alpha > 0 .$ This results in that
the dynamo waves should propagate to the pole in contradiction to
the solar magnetic field observations (see, {\em e.g.,}
\cite{M78,P79,KR80,ZRS83,S89}).

In this study we found that in a developed turbulent
convection the $\alpha$-effect can change its sign depending on the
rate of rotation and an anisotropy of turbulence. In the lower part of the
solar convective zone the fluid rotation is very fast
in comparison with the turnover time
of turbulent eddies. In this region $ \alpha > 0 .$ In the upper part
of the solar convective zone the fluid rotation is very slow and
$ \alpha < 0 .$ This explains the observed properties of the solar dynamo
waves. The growth of the mean magnetic field is saturated by nonlinear effects
(see, {\em e.g.,} \cite{KRR94,F99,KR99,RK2000,KMRS2000}). The $ 22 $-years
solar magnetic activity is also poorly understood. A characteristic time
of the turbulent magnetic diffusion in the solar convective zone is of
the order 2-3 years and it cannot explain the characteristic time of
solar magnetic activity. We found that the fast rotation causes an
additional effective drift velocity of a mean magnetic field
that can increase the period of the dynamo waves provides
the $ 22 $-years solar magnetic activity.

\section{The governing equations and the method of the derivations}

Our goal is to study an effect of rotation on a developed
turbulent stratified convection. This allows us to derive
dependencies of the hydrodynamic helicity, the alpha-effect and
the effective drift velocity of the mean magnetic field on the
angular velocity. To this end we consider a fully developed
turbulent convection in a stratified rotating fluid with large
Rayleigh and Reynolds numbers. The governing equations are given
by
\begin{eqnarray}
{D {\bf u} \over D t} &=& - \bec{\nabla} \biggl({P \over
\rho_{_{0}} } \biggr) + 2 {\bf u} \times {\bf \Omega} - {\bf g} S
+ {\bf f}_{\nu} \;,
\label{B1} \\
{D S \over D t} &=& - {\bf u} \cdot {\bf N}_{b} - {1 \over T_{0}}
\, \bec{\nabla} \cdot {\bf F}_{\kappa}(S) \;, \label{B2}
\end{eqnarray}
where $ {\bf u} $ is the fluid velocity with $\bec{\nabla} \cdot
{\bf u} = {\bf \Lambda} \cdot {\bf u} ,$ $ \quad D / Dt = \partial
/ \partial t + {\bf u} \cdot \bec{\nabla} ,$ $ \quad {\bf \Omega}
$ is the angular velocity, $ {\bf g} $ is the gravity field that
includes an effect of the centrifugal force, $ \rho_{_{0}} {\bf
f}_{\nu} $ is the viscous force, $ {\bf F}_{\kappa}(S) $ is the
thermal flux that is associated with the molecular thermal
conductivity, $ {\bf \Lambda} = - \rho_{_{0}}^{-1} \bec{\nabla}
\rho_{_{0}} ,$ $ \quad {\bf N}_{b} = (\gamma P_{0})^{-1}
\bec{\nabla} P_{0} - \rho_{_{0}}^{-1} \bec{\nabla} \rho_{_{0}} .$
The variables with the subscript $ "0" $ corresponds to the
hydrostatic equilibrium (i.e., the hydrostatic basic reference
state):
\begin{eqnarray}
\bec{\nabla} P_{0} = \rho_{_{0}} {\bf g} \;, \label{K20}
\end{eqnarray}
and $ T_{0} $ is the equilibrium fluid temperature, $ S = P /
\gamma P_{0} -  \rho / \rho_{_{0}} $ are the deviations of the
entropy from the hydrostatic equilibrium, $ P $ and $ \rho $ are
the deviations of the fluid pressure and density from the
hydrostatic equilibrium. The Brunt-V\"{a}is\"{a}l\"{a} frequency,
$ \tilde \Omega_{b} ,$ is determined by the equation $ \tilde
\Omega_{b}^{2} = - {\bf g} \cdot {\bf N}_{b} .$ To derive
Eq.~(\ref{B1}) we use an identity: $ - \bec{\nabla} P + {\bf g}
\rho = - \rho_0 [\bec{\nabla} (P / \rho_0) + {\bf g} S - P {\bf
N}_{b}/ \rho_0] ,$ where we assumed that $ |P {\bf N}_{b}/ \rho_0|
\ll |{\bf g} S| ,$ $ \, |P {\bf N}_{b}/ \rho_0| \ll |\bec{\nabla}
(P / \rho_0)| .$ This assumption corresponds to nearly isentropic
basic reference state when $ {\bf N}_{b} $ is very small. For the
derivation of this identity we also used Eq.~(\ref{K20}). We also
consider a low-Mach-number fluid flow with a very small frequency
$ \tilde \Omega_{b} ,$ {\em i.e.,} $ | \tilde \Omega_{b} | \ll
\sqrt{g \Lambda} $ and $ | \tilde \Omega_{b} \tau |^{2} \ll 1 ,$
where $ \tau $ is the correlation time of the turbulent velocity
field. Equations~(\ref{B1}) and (\ref{B2}) are written in the
Boussinesq approximation for $ \bec{\nabla} \cdot {\bf u} \not= 0
.$ This is more usually called "the anelastic approximation".

Now we consider a purely hydrostatic isentropic basic reference
state, i.e., $ \tilde\Omega_{b} = 0 $ $ \, ({\bf N}_{b} = 0) .$
Thus the turbulent convection is regarded as a small deviation
from a well-mixed adiabatic state (for more discussion, see
\cite{BR95}). We will use a mean field approach whereby the
velocity, pressure and entropy are separated into the mean and
fluctuating parts. Using Eqs.~(\ref{B1}) and (\ref{B2}) we derive
equations for the turbulent fields: $ v_{z} = \sqrt{\rho_0(z)} \,
u_{z} ,$ $ w = \sqrt{\rho_0(z)} \, (\bec{\nabla} {\bf \times} {\bf
u})_{z} $ and $ s = \sqrt{\rho_0(z)} \, (S - \bar S) ,$ where $
\bar S \equiv \langle S \rangle $ is the mean entropy, the angular
brackets denote ensemble averaging, and for simplicity we consider
turbulent flow with zero mean velocity. Here $ \rho_0(z) $ is
dimensionless density measured in the units of $ \rho_0(z=0). $
The equations for the turbulent fields are given by
\begin{eqnarray}
\biggl({\Lambda^{2} \over 4} - \Delta \biggr) {\partial v_{z} \over
\partial t} &=& (2 {\bf \Omega} \cdot \bec{\nabla} +
{\bf \Omega} \cdot {\bf \Lambda}) w + 2 \Lambda \Omega_{x}
{\partial v_{z} \over \partial y}
\nonumber\\
& & - g \Delta_{\perp} s + V_{N} \;,
\label{B4} \\
{\partial w \over \partial t} &=& (2 {\bf \Omega} \cdot \bec{\nabla} -
{\bf \Omega} \cdot {\bf \Lambda}) v_{z} + W_{N} \;,
\label{B5} \\
{\partial s \over \partial t} &=& - {\Omega_{b}^{2} \over g} v_{z}
+ S_{N} \;, \label{B6}
\end{eqnarray}
where $ \Omega_{b}^{2} = - {\bf g} \cdot \bec{\nabla} \bar S ,$ $
\, \Delta_{\perp} = \Delta - \partial^{2} / \partial z^{2} ,$ $
\quad V_{N} ,$ $ W_{N} $ and $ S_{N} $ are the nonlinear terms
which include the molecular dissipative terms [see Eqs.
(\ref{RD1})-(\ref{RD3}) in Appendix A], the field $ {\bf g} $ is
directed opposite to the axis $ z $ and $ {\bf \Omega} =
(\Omega_{x}, 0, \Omega_{z}) .$ We assumed here that $ \Lambda^{-1}
| \partial \Lambda / \partial z | \ll \Lambda .$
Equation~(\ref{B4}) follows from Eq.~(\ref{B1}) after the
calculation $ [\bec{\nabla} {\bf \times} (\bec{\nabla} {\bf
\times} {\bf u})]_{z} .$

By means of Eqs. (\ref{B4})-(\ref{B6}) we derive  dependencies of
the hydrodynamic helicity, the alpha-effect and the effective
drift velocity on the angular velocity. The procedure of the
derivation is outlined in the following (for details, see
Appendixes A, B and C).

(a). Using Eqs. (\ref{B4})-(\ref{B6}) we derive equations for the following
second moments:
\begin{eqnarray*}
f_{ij}({\bf k}) &=& \hat L(v_{i},v_{j}) \;, \quad \chi({\bf k}) =
\hat L(w,v_{z}) \;,
\\
F({\bf k}) &=& \hat L(s,w) \;, \quad G({\bf k}) = \hat L(w,w) \;,
\\
\Phi_i({\bf k}) &=& \hat L(s,v_{i}) \;, \quad \Theta({\bf k}) =
\hat L(s,s) \;,
\end{eqnarray*}
where $ \hat L(a,b) = \langle a({\bf k}) b(-{\bf k}) \rangle $ and
$ {\bf v} = \sqrt{\rho_0(z)} \, {\bf u} .$ The equations for these
correlation functions are given by Eqs. (\ref{B13})-(\ref{B18}) in
Appendix A. In this derivation we assumed that $ \Lambda^{2} \ll
k^{2} .$

(b). The equations for the second moments contain third moments
and a problem of closing the equations for the higher moments
arises. Various approximate methods have been proposed for the
solution of problems of this type (see, {\em e.g.,}
\cite{MY75,O70,Mc90}). The simplest procedure is the $ \tau $
approximation, which is widely used in the theory of kinetic
equations. For magnetohydrodynamic turbulence this approximation
was used in \cite{PFL76} (see also \cite{RK2000,KRR90,KMR96}). One
of  the simplest procedures which allows us to express the third
moments $ f_{_{N}} ,$ $ \chi_{_{N}} , \ldots ,$ $ \Theta_{N} $ in
Eqs. (\ref{B13})-(\ref{B18}) in terms of the second moments, reads
\begin{eqnarray}
f_{_{N}}({\bf k}) - f_{_{N}}^{(0)}({\bf k}) &=& - {f({\bf k}) -
f^{(0)}({\bf k}) \over \tau (k)} \;, \label{B19}
\end{eqnarray}
and similarly for other third moments, where $ f({\bf k}) = e_i
e_j f_{ij}({\bf k}) ,$ $ \, {\bf e}$ is the unit vector directed
along the axis $z ,$ the superscript $ (0) $ corresponds to the
background turbulent convection (it is a turbulent convection
without rotation, $ {\bf \Omega} = 0),$ and $ \tau (k) $ is the
characteristic relaxation time of the statistical moments. Note
that we applied the $ \tau $-approximation only to study the
deviations from the background turbulent convection which is
caused by the rotation. The background turbulent convection is
assumed to be known.

The $ \tau $-approximation  is in general similar to Eddy Damped
Quasi Normal Markowian (EDQNM) approximation. However, there is a
principle difference between these two approaches (see
\cite{O70,Mc90}). The EDQNM closures do not relax to equilibrium,
and this procedure does not describe properly the motions in the
equilibrium state in contrast to the $ \tau $-approximation.
Within the EDQNM theory, there is no dynamically determined
relaxation time, and no slightly perturbed steady state can be
approached \cite{O70}. In the $ \tau $-approximation, the
relaxation time for small departures from equilibrium is
determined by the random motions in the  equilibrium state, but
not by the departure from equilibrium \cite{O70}. As follows from
the analysis performed in \cite{O70} the $ \tau $-approximation
describes the relaxation to the equilibrium state (the background
turbulent convection) more accurately than the EDQNM approach.

Note that we analyzed the applicability of the
$\tau$-approximation for description of the mean-field dynamics of
the mean magnetic field and mean scalar fields by comparison of
the derived mean-field equations using other methods such as the
path-integral approach and the renormalization group approach (see
\cite{KR94,RK97,EKR96,EKR97,EKRP97}. This comparison showed that
the $\tau$-approximation yields the results similar to that
obtained by means of the other methods.

(c). We assume that the characteristic times of variation of the
second moments $ f({\bf k}) ,$ $ \chi({\bf k}) , \ldots ,$ $
\Theta({\bf k}) $ are substantially larger than the correlation
time $ \tau(k) $ for all turbulence scales. This allows us to
determine a stationary solution for the second moments $ f({\bf
k}) ,$ $ \chi({\bf k}) , \ldots ,$ $ \Theta({\bf k}) $ [see Eqs.
(\ref{B20})-(\ref{B28}) in Appendix A].

(d). For the integration in $ {\bf k} $-space of the second
moments $ f({\bf k}) ,$ $ \chi({\bf k}) , \ldots ,$ $ \Theta({\bf
k}) $ we have to specify a model for the background turbulent
convection (without rotation). Here we use the following model of
the background turbulent convection which will be discussed in
more details in Appendix D:
\begin{eqnarray}
f_{ij}^{(0)}({\bf k}) &=& f_{\ast} \tilde W(k) [P_{ij}({\bf k}) +
\varepsilon P_{ij}^{(\perp)}({\bf k}_{\perp})] \;,
\label{CB8} \\
\Phi^{(0)}_{i}({\bf k}) &=& k_{\perp}^{-2} [k^2
\Phi_{z}^{(0)}({\bf k}) e_{j} P_{ij}({\bf k})
\nonumber\\
& & + i F^{(0)}({\bf k}) ({\bf e} {\bf \times} {\bf k})_{i}] \;,
\label{CB9} \\
\Phi_{z}^{(0)}({\bf k}) &=& \Phi^{\ast}_z \tilde W(k) \biggl[2
\sigma - 3 (\sigma - 1)  \biggl({k_{\perp} \over k} \biggr)^{2}
\biggr] \;,
\label{CB10} \\
F^{(0)}({\bf k}) &=& - 6 i ({\bf \Phi}^{\ast} \cdot ({\bf e} {\bf
\times} {\bf k})) f^{(0)}({\bf k}) / f_{\ast} \;,
\label{CB11} \\
G^{(0)}({\bf k}) &=& (1 + \varepsilon) f^{(0)}({\bf k}) k^{2} \;,
\label{CCB12}\\
\Theta^{(0)}({\bf k}) &=&  2 \Theta_{\ast} \tilde W(k) \;,
\label{CB12}
\end{eqnarray}
where $ f_{ij}({\bf k}) = \langle v_i({\bf k}) v_j (-{\bf k})
\rangle ,$ $\, \tilde W(k) = W(k) / 8 \pi k^{2} ,$ $ \,
f^{(0)}({\bf k}) = f_{\ast} (k_{\perp} / k)^{2} \tilde W(k) ,$ and
$ f_{ij}^{(0)}({\bf k}) e_{ij} = f^{(0)}({\bf k}) ,$
\begin{eqnarray*}
\varepsilon = {2 \over 3} \biggl( {\langle {\bf u}_{\perp}^{2}
\rangle \over \langle {\bf u}_{z}^{2} \rangle} - 2\biggr)
\end{eqnarray*}
is the degree of anisotropy of the turbulent velocity field $ {\bf
u} = {\bf u}_{\perp} + u_{z} {\bf e} .$ Here $ P_{ij}({\bf k}) =
\delta_{ij} - k_{ij} ,$ $ \, k_{ij} = k_{i} k_{j} / k^{2} ,$ $ \,
{\bf k} = {\bf k}_{\perp} + k_{z} {\bf e} ,$ $ \, k_{z} = {\bf k}
\cdot {\bf e} ,$ $ \, P_{ij}^{(\perp)}({\bf k}_{\perp}) =
\delta_{ij} - k^{\perp}_{ij} - e_{ij} ,$ $ \, k^{\perp}_{ij} =
({\bf k}_{\perp})_{i} ({\bf k}_{\perp})_{j} / k_{\perp}^{2} ,$ $
\, e_{ij} = e_{i} e_{j} ,$ $ \, \sigma $ is the degree of
anisotropy of the turbulent flux of entropy (see below). We assume
that $ \tau(k) = 2 \tau_{_{0}} \bar \tau(k) ,$ $ \, W(k) = - d
\bar \tau(k) / dk ,$ $ \, \bar \tau(k) = (k / k_{0})^{1-q} ,$ $\,
1 < q < 3 $ is the exponent of the kinetic energy spectrum (e.g.,
$ q = 5/3 $ for Kolmogorov spectrum), $ k_{0} = 1 / l_{0} ,$ and $
l_{0} $ is the maximum scale of turbulent motions, $ \tau_{_{0}} =
l_{0} / u_{0} $ and $ u_{0} $ is the characteristic turbulent
velocity in the scale $ l_{0} .$ Motion in the background
turbulent convection is assumed to be non-helical. In
Eqs.~(\ref{CB8}) and~(\ref{CB9}) we neglected small terms $ \sim
O(\Lambda f_{\ast}) $ and $ \sim O(\Lambda \Phi^{\ast}_z) ,$
respectively. Now we calculate $ f_{ij}^{(0)} \equiv \int
f_{ij}^{(0)}({\bf k}) \,d {\bf k} $ using Eq.~(\ref{CB8}): $
f_{ij}^{(0)} = (f_{\ast} / 3) [\delta_{ij} + (3 \varepsilon / 4)
(\delta_{ij} - e_{ij})] .$ Note that $ - 4/3 \leq \varepsilon <
\infty .$ The lower limit of $ \varepsilon $ follows from the
condition $ f_{xx}^{(0)} \geq 0 $ (or $ f_{yy}^{(0)} \geq 0) .$
Similarly, using Eqs. (\ref{CB9})-(\ref{CB11}) we obtain $ {\bf
\Phi}^{(0)} \equiv \int {\bf \Phi}^{(0)}({\bf k}) \,d {\bf k} =
{\bf \Phi}^{\ast} .$ The parameter $ \sigma $ can be presented in
the form
\begin{eqnarray}
\sigma &=& {1 + \tilde \xi (q + 1) / (q - 1) \over 1 + \tilde \xi
/ 3} \;,
\label{CD6} \\
\tilde \xi  &=& (l_{\perp} / l_{z})^{q-1} - 1 \;, \label{CD7}
\end{eqnarray}
where $ l_{\perp} $ and $ l_{z} $ are the horizontal and vertical
scales in which the two-point correlation function $
\Phi_{z}^{(0)}({\bf r}) = \langle s({\bf x}) {\bf u}({\bf x}+{\bf
r}) \rangle $ tends to zero. The parameter $ \tilde\xi $
determines the degree of thermal anisotropy. In particular, when $
l_{\perp} = l_{z} $ the parameter $ \tilde\xi = 0 $ and $ \sigma =
1 .$ For $ l_{\perp} \ll l_{z} $ the parameter $ \tilde\xi = - 1 $
and $ \sigma = - 3 / (q-1) .$ The maximum value $ \tilde \xi_{\rm
max} $ of the parameter $ \tilde \xi $ is given by $ \tilde
\xi_{\rm max} = q - 1 $ for $ \sigma = 3 .$ Thus, for $ \sigma < 1
$ the thermal structures have the form of column or thermal jets $
(l_{\perp} < l_{z}) ,$ and for $ \sigma > 1 $ there exist the
`'pancake'' thermal structures $ (l_{\perp} > l_{z}) $ in the
background turbulent convection.

The relationship between $ \Phi^{\ast}_z $ and $ f_{\ast} $
follows from Eq.~(\ref{P2}) for the kinetic turbulent energy $
\rho_{_{0}} \langle {\bf u}^{2} \rangle ,$ and it is given by $
f_{\ast} = 2 \lambda g \tau_{_{0}} \Phi^{\ast}_z / \varepsilon ,$
where $ \lambda = 2 \varepsilon \delta_{\ast} / (\varepsilon + 2)
$ and $ \delta_{\ast} = (3-q) / 2(q-1) .$ Note that for Kolmogorov
spectrum $ q = 5/3 $ and $ \delta_{\ast} = 1 .$ In Section III we
will present results for $ \delta_{\ast} = 1 .$ For the
integration in $ {\bf k} $-space we used identities given in
Appendixes B and C.

Thus, the ''input parameters" in the theory include the parameters
that describe the model of background turbulent convection, i.e.,
the degree of anisotropy of the turbulent velocity field $
\varepsilon ,$ the degree of anisotropy of the turbulent flux of
entropy $ \sigma ,$ the maximum scale of turbulent motions $ l_0
,$ the turbulent velocity $ u_0 \equiv \sqrt{\langle {\bf u}^2
\rangle} = \sqrt{f_{pp}^{(0)}} $ (the r.m.s. velocity) in the
maximum scale of turbulent motions, the exponent of the kinetic
energy spectrum $ q .$ The ''input parameters" also include the
density stratification length $ \Lambda $ and the angular velocity
$ {\bf \Omega} .$ Note that $ f_{\ast} = u_0^2 / (1 + \varepsilon
/ 2) $ and $ \Phi^{\ast}_z = u_0^2 / (2 \delta_{\ast} g
\tau_{_{0}}) .$ The described above procedure allows us to
determine the dependencies of the hydrodynamic helicity, the
alpha-effect and the effective drift velocity of the mean magnetic
field on the rate of rotation.

The considered model of a background turbulent convection written
in ${\bf k}$-space is enough general and it does not contradict to
the known Nusselt number dependencies on Rayleigh number. On the
other hand, the observations of the turbulent convection on the
surface of the Sun cannot give the Nusselt number dependence on
Rayleigh number, i.e., it is possible to obtain only one point in
this curve. The parameters $ \varepsilon ,$ $ \, u_0 ,$ $ \, l_0
,$ $ {\bf \Omega} ,$ etc can be calculated from the solar
observations. In addition, the direct numerical simulations of
turbulent convection (see \cite{BN90,OSB01,OSB02}) are in an
agreement with our model of turbulent convection.

\section{Effect of rotation}

In this Section we present the results of the calculations (described
above) for the hydrodynamic helicity, the alpha-effect
and the effective drift velocity of the mean magnetic field
as the functions of the rate of rotation and an anisotropy of turbulence.

\subsection{The hydrodynamic helicity}

Using Eqs.~(\ref{D2}) and (\ref{D8}) in Appendix A we find the
dependence of the hydrodynamic helicity $ \chi^{(v)} = \langle
{\bf u}~\cdot~(\bec{\nabla} {\bf \times} {\bf u}) \rangle $ on the
angular velocity:
\begin{eqnarray}
\chi^{(v)} &=& - {1 \over 12} \biggl({l_{0}^{2} \Omega \over
L_{\rho} \tau_{_{0}}} \biggr) \, [\Psi_{1}(\omega) +
\Psi_{2}(\omega) \sin^{2} \phi_l
\nonumber\\
& & + \Psi_{3}(\omega) \sin^{4} \phi_l] \sin \phi_l \;
\label{D10}
\end{eqnarray}
(for details, see Appendix A), where $ \omega = 4 \tau_{_{0}}
\Omega ,$ $ \, l_{0} = u_{0} \tau_{_{0}} ,$ $ \,  u_{0}^{2} = 2 g
\tau_{_{0}} \Phi^{\ast}_z \delta_\ast , $ $ \, \sin \phi_l =
\bec{\hat \omega} \cdot {\bf e} ,$ $\, \phi_l $ is the latitude, $
\, \bec{\hat \omega} = {\bf \Omega} / \Omega ,$ $ \, {\bf e} $ is
the unit vector directed along the $z$-axis, $ \, L_{\rho} =
\Lambda^{-1} ,$ the functions $ \Psi_{m}(\omega) $ are given by
Eqs. (\ref{M50}) in Appendix C. Hereafter we assume that
$\delta_\ast=1 .$ For a slow rotation $ (\omega \ll 1) $ the
hydrodynamic helicity $ \chi^{(v)} $ is given by
\begin{eqnarray}
\chi^{(v)} \approx - {1 \over 6} \biggl({l_{0}^{2} \Omega \over
L_{\rho} \tau_{_{0}}} \biggr) \sin \phi_l \left({164 \sigma \over
15} + {12 \over 5} - 5 \lambda \right) \;, \label{D11}
\end{eqnarray}
and for $ \omega \gg 1 $ it is given by
\begin{eqnarray}
\chi^{(v)} \approx {3 \pi \over 8} \biggl({l_{0} u_{0} \over
L_{\rho} \tau_{_{0}}} \biggr) \lambda \biggl(1 + {1 \over 4}
\sin^{2} \phi_l \biggr) \sin \phi_l \; . \label{D12}
\end{eqnarray}
Note that the meaning $\omega \equiv 4 \Omega \tau_{_{0}} \gg 1$
is $\omega$ large, but only up to some upper limit, i.e., an
intermediate range of values. This implies that the rotation
cannot be very fast to affect the correlation time $\tau(k)$ of
turbulent velocity field in its inertial range. Also we assumed
that the parameters $ \varepsilon $ and $ \sigma $ are independent
of $\omega .$

\subsection{The $ \alpha $-effect}

Now we find the dependence of the $ \alpha $-effect on the angular
velocity. To this end we use the induction equation for the magnetic
field
\begin{eqnarray}
{\partial {\bf H} \over \partial t} = \bec{\nabla} \times ({\bf u}
\times {\bf H} - \eta \bec{\nabla} \times {\bf H}) \;, \label{F1}
\end{eqnarray}
where $ \eta $ is the magnetic diffusion due to the electrical
conductivity of fluid. The magnetic field, ${\bf H}$, is divided
into the mean and fluctuating parts: $ {\bf H} = {\bf B} + {\bf b}
,$ where the mean magnetic field $ {\bf B} = \langle {\bf H}
\rangle $ and $ {\bf b} $ is the fluctuating field. An equation
for $ {\bf h} = \sqrt{\rho_0} \, {\bf b} $ follows from
Eq.~(\ref{F1}) and it is given by
\begin{eqnarray}
{\partial {\bf h} \over \partial t} &=& ({\bf B} \cdot
\bec{\nabla}){\bf v} - ({\bf v} \cdot \bec{\nabla}){\bf B} - ({\bf
v} \cdot {\bf \Lambda}){\bf B}
\nonumber \\
& & + {1 \over 2} ({\bf B} \cdot {\bf \Lambda}){\bf v} + {\bf
H}_{N} \;, \label{F2}
\end{eqnarray}
where $ {\bf H}_{N} $ are the nonlinear terms which also include
the magnetic diffusion term [see Eq. (\ref{RD4}) in Appendix A].
In order to derive equation for the $ \alpha $-tensor we introduce
the electromotive force $ {\cal E}_{i} = \langle {\bf u} \times
{\bf b} \rangle_{i} = \rho_0^{-1} \varepsilon_{imn} \int
\chi^{(c)}_{mn}({\bf k}) \,d {\bf k} ,$ where $
\chi^{(c)}_{ij}({\bf k}) = \langle v_{i}({\bf k}) h_{j}(-{\bf k})
\rangle \equiv \hat L(v_i, h_j) .$ A general form of the
electromotive force is given by $ {\cal E}_{i} = \alpha_{ij} B_{j}
+ ({\bf V}^{\rm eff} {\bf \times} {\bf B})_{i} - \eta_{ij}
(\bec{\nabla} {\bf \times} {\bf B})_{j} - \kappa_{ijk} (\partial
\hat B)_{ij} - [\bec{\delta} {\bf \times} (\bec{\nabla} {\bf
\times} {\bf B})]_{i} = a_{ij} B_{j} + b_{ijk} B_{i,j} $ (see,
{\em e.g.,} \cite{R80} and Appendix A), where the tensors $
\alpha_{ij} $ and $ \eta_{ij} $ describe the $ \alpha $-effect and
turbulent magnetic diffusion, respectively, $ {\bf V}^{\rm eff} $
is the effective diamagnetic (or paramagnetic) velocity, $
\kappa_{ijk} $ and $ \bec{\delta} $ describe a nontrivial behavior
of the mean magnetic field in an anisotropic turbulence, $ B_{i,j}
= \nabla_{j} B_{i} $ and $ (\partial \hat B)_{ij} = (1/2) (B_{i,j}
+ B_{j,i}) .$ The $ \alpha $-tensor, $ \alpha_{ij} ,$ is
determined by a symmetric part of the tensor $ a_{ij} ,$ {\rm
i.e.,} by $ a_{ij}^{(S)} \equiv (1/2)(a_{ij} + a_{ji}) .$ The
tensor $ a_{ij} $ is calculated in Appendix A. The $ \alpha
$-tensor is given by
\begin{eqnarray}
\alpha_{ij} &=& {1 \over 6} \biggl({l_{0}^{2} \Omega \over
L_{\rho}} \biggr) \{ \sin \phi_l \, [ (\Psi_{4}(\omega) +
\Psi_{5}(\omega) \sin^{2} \phi_l) \, \delta_{ij}
\nonumber \\
& & + (\Psi_{6}(\omega) +  \Psi_{7}(\omega) \sin^2 \phi_l) \,
\omega_{ij} + \, \Psi_{8}(\omega) \, e_{ij}]
\nonumber \\
& & + [\Psi_{9}(\omega) + \Psi_{10}(\omega) \sin^2 \phi_l] \,
(e_{i} \hat \omega_{j} + e_{j} \hat \omega_{i}) \} \;,
\label{F12}
\end{eqnarray}
(for details, see Appendix A), where $ \omega_{ij} = \hat
\omega_{i} \hat \omega_{j} ,$ $ \, e_{ij} = e_{i} e_{j} ,$ the
functions $ \Psi_{m}(\omega) $ are given  by Eqs. (\ref{M50}) in
Appendix C. Here we present asymptotic formulas for the isotropic
part $ (\alpha_{ij}^{({\rm isotr})} \equiv \alpha \delta_{ij}) $
of the $ \alpha $-tensor. For a slow rotation $ (\omega \ll 1) $
the parameter $ \alpha $ is given by
\begin{eqnarray}
\alpha \approx {4 \over 5} \biggl({l_{0}^{2} \Omega \over
L_{\rho}} \biggr) \biggl(2 - {\sigma \over 3} - {5 \lambda \over
6} \biggr) \sin \phi_l \;, \label{F14}
\end{eqnarray}
and for $ \omega \gg 1 $ it is given by
\begin{eqnarray}
\alpha & \approx & - {\pi \over 32} \biggl({l_{0} u_{0} \over
L_{\rho}} \biggr) \biggl(2 \lambda + {\sigma \over 3} - 3
\nonumber \\
& & + (\sigma - 1) \sin^2 \phi_l \biggr) \sin \phi_l \; .
\label{F15}
\end{eqnarray}
It is seen from Eqs.~(\ref{D11}) and~(\ref{F14}) that for a slow
rotation and isotropic background turbulent convection $ (\sigma =
1 $ and $ \varepsilon = 0), $ the parameter $ \alpha \approx -
(5/27) \tau_{_{0}} \chi^{(v)} ,$ where $ \chi^{(v)} = \langle {\bf
u} \cdot (\bec{\nabla} {\bf \times} {\bf u}) \rangle .$ However,
when a rotation is not slow, the latter relationship does not
valid.

\begin{figure}
\centering
\includegraphics[width=8cm]{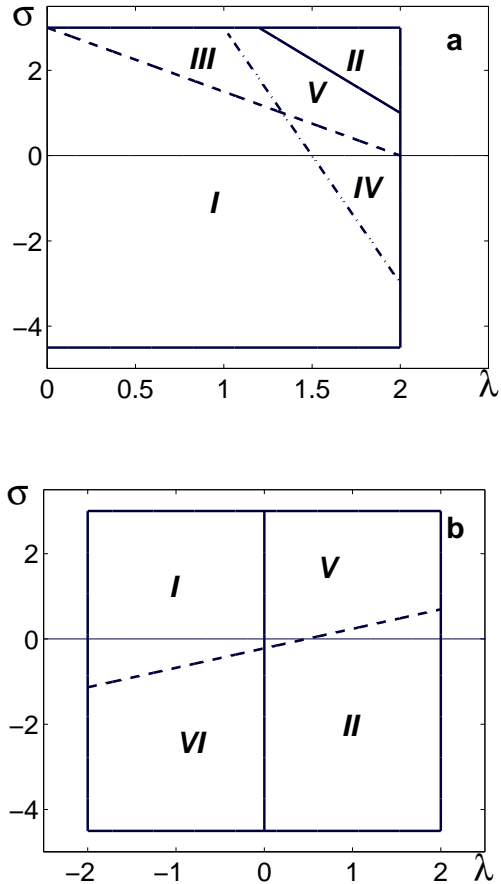}
\caption{\label{Fig1} Characteristic ranges of parameters with
different behavior of $\alpha$-effect (FIG.1a) and the parameter $
\alpha_{\chi} = - (1/3) \tau_{_{0}} \chi^{(v)} $ (FIG.1b). The
range I for the $\alpha$-effect (FIG.1a) also exists for $ - 2 <
\lambda < 0 $ and $ - 9/2 < \sigma < 3 .$}
\end{figure}

The $\alpha$-effect depends on the degrees of the velocity
anisotropy $ \varepsilon $ and the thermal anisotropy $ \sigma .$
Asymptotic formulas for a slow rotation $ (\omega \ll 1) $ and for
$ \omega \gg 1 $ show that there are several characteristic ranges
of parameters with different behavior of $\alpha$-effect. In
FIG.~1a these ranges are separated by lines $ \sigma = 3(3 - 2
\lambda) ,$ $ \, \sigma = 3(1 - \lambda / 2) $ and $ \sigma = 6 -
5 \lambda / 2 ,$ where $ - 2 < \lambda < 2 ,$ $\, - 9/2 < \sigma <
3 ,$ and $ 0 \leq \phi_l \leq \pi / 2 .$ Here $ \lambda = 2
\varepsilon / (\varepsilon + 2) .$ In the ranges I and II the
$\alpha$-effect does not change its sign for all $ \Omega
\tau_{_{0}} $ and $ \phi_l . $ In particular, in the range I:
$\alpha > 0 $ and in the range II: $\alpha < 0 .$ In the range V
the $\alpha$-effect changes its sign at a certain value of $
\Omega \tau_{_{0}} $ for all $ \phi_l .$ In the ranges III and IV
the $\alpha$-effect changes its sign at a certain  value of $
\Omega \tau_{_{0}} $ and a certain range of the latitudes  $
\phi_l .$ In the range III the degree of thermal anisotropy $
\sigma > 1 $ (which corresponds to the ''pancake" small-scale
thermal structure of the background turbulent convection), and in
the range IV the degree of thermal anisotropy $ \sigma < 1 $
(i.e., a column-like thermal structure). The $\alpha$-effect can
be negative for a slow rotation only in the range II. Note that
the negative $\alpha$-effect corresponds to the propagation of the
solar dynamo waves to the equator.

Our analysis shows that when the rotation is not slow, the
$\alpha$-effect is determined not only by the contributions from
the hydrodynamic helicity and its behavior is much more
complicated in a rotating fluid. In order to demonstrate this we
plotted in FIGS. 2-4 the dependencies of the $\alpha$-effect
(solid line) and $ \alpha_{\chi} \equiv - (1/3) \tau_{_{0}}
\chi^{(v)} $ (dashed line) on the parameter $ \Omega \tau_{_{0}} $
for different latitudes (FIG. 2 is for the latitude $ \phi_{l} =
15^{\circ} ,$ FIG. 3 is for $ \phi_{l} = 35^{\circ} $ and FIG. 4
is for $ \phi_{l} = 90^{\circ}).$ Here the parameters $\alpha$ and
$ \alpha_{\chi} $ are measured in the units of $ l_{0} u_{0} / 4
L_{\rho} .$ Figures 2-4 demonstrate that the functions
$\alpha(\Omega \tau_{_{0}})$ and $ \alpha_{\chi}(\Omega
\tau_{_{0}}) $ are totally different. For example, in the case
$\varepsilon = 13$ and $\sigma = 0 $ the $\alpha$-effect and $
\alpha_{\chi} $ have opposite signs for all $ \Omega \tau_{_{0}} $
(see FIG. 4c).

\begin{figure}
\centering
\includegraphics[width=8cm]{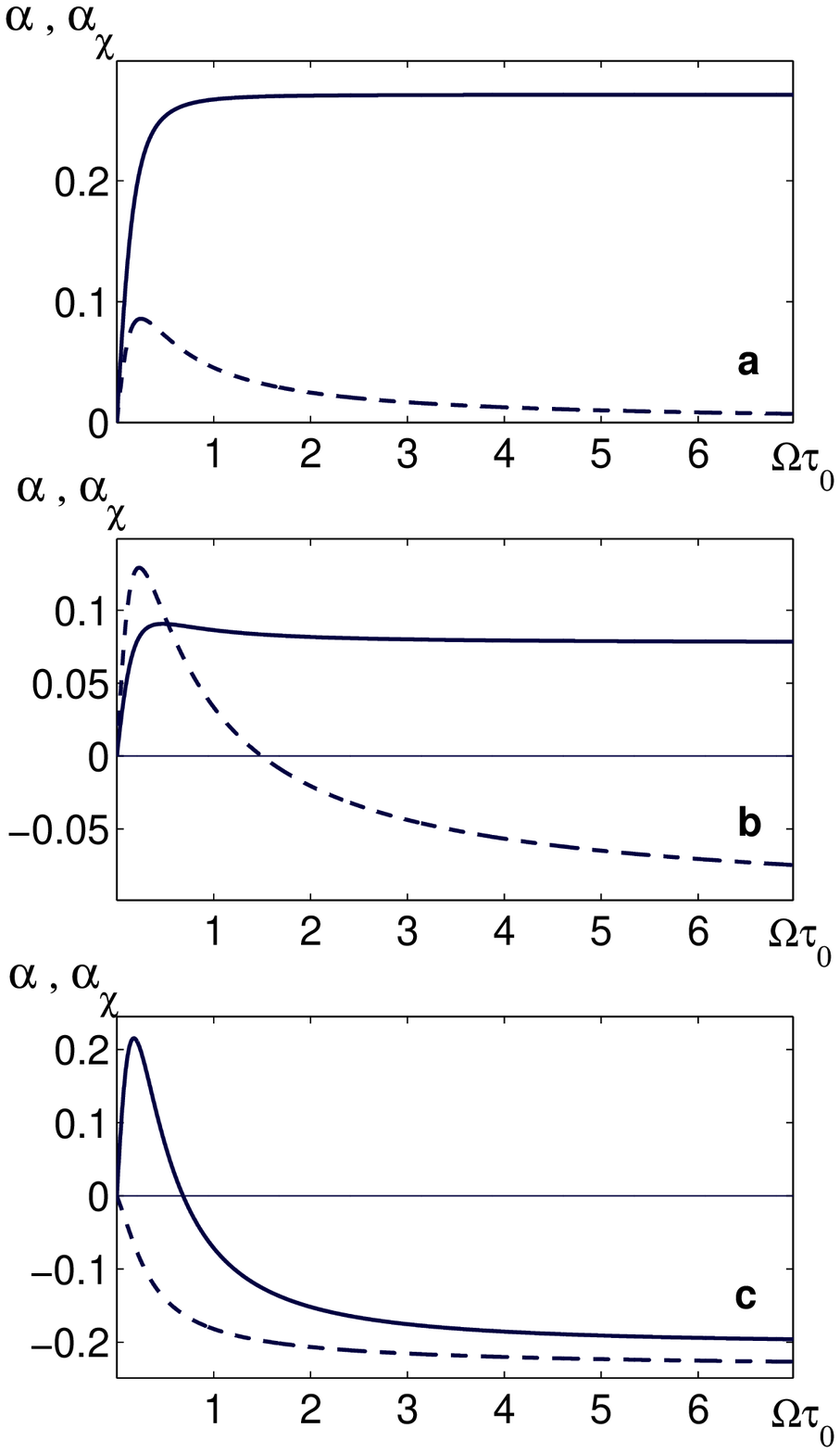}
\caption{\label{Fig2} The $\alpha$ (solid line) and $
\alpha_{\chi} = - (1/3) \tau_{_{0}} \chi^{(v)} $ (dashed line) as
the functions of the parameter $ \Omega \tau_{_{0}} $ for $
\phi_{l} = 15^{\circ} $ and different values of the degrees of
anisotropy: (a). $\varepsilon = 0$ and $\sigma = 1$; (b).
$\varepsilon = 1.2 $ and $\sigma = 2$; (c). $\varepsilon = 13$ and
$\sigma = 0 .$ In FIG. 2c the $\alpha$ effect is multiplied by
$ 5 .$}
\end{figure}

\begin{figure}
\centering
\includegraphics[width=8cm]{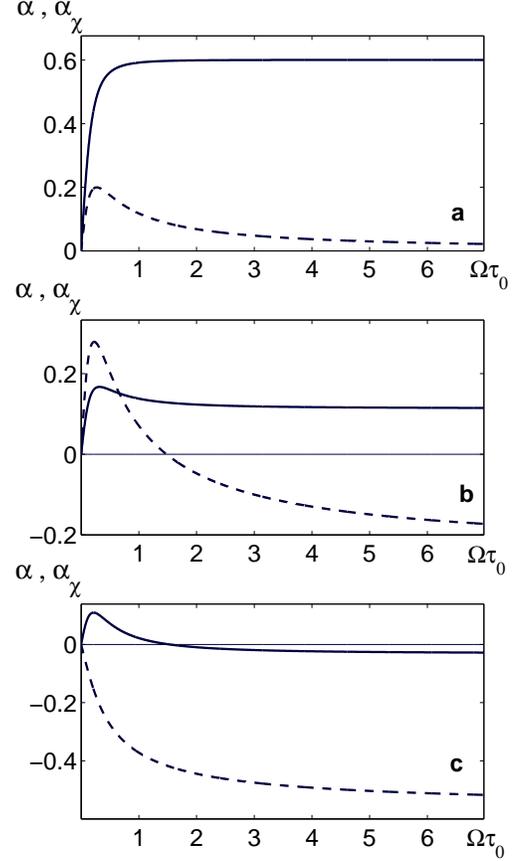}
\caption{\label{Fig3} The $\alpha$ (solid line) and $
\alpha_{\chi} = - (1/3) \tau_{_{0}} \chi^{(v)} $ (dashed line) as
the functions of the parameter $ \Omega \tau_{_{0}} $ for $
\phi_{l} = 35^{\circ} $ and different values of the degrees of
anisotropy: (a). $\varepsilon = 0$ and $\sigma = 1$; (b).
$\varepsilon = 1.2 $ and $\sigma = 2$; (c). $\varepsilon = 13$ and
$\sigma = 0 .$}
\end{figure}

\begin{figure}
\centering
\includegraphics[width=8cm]{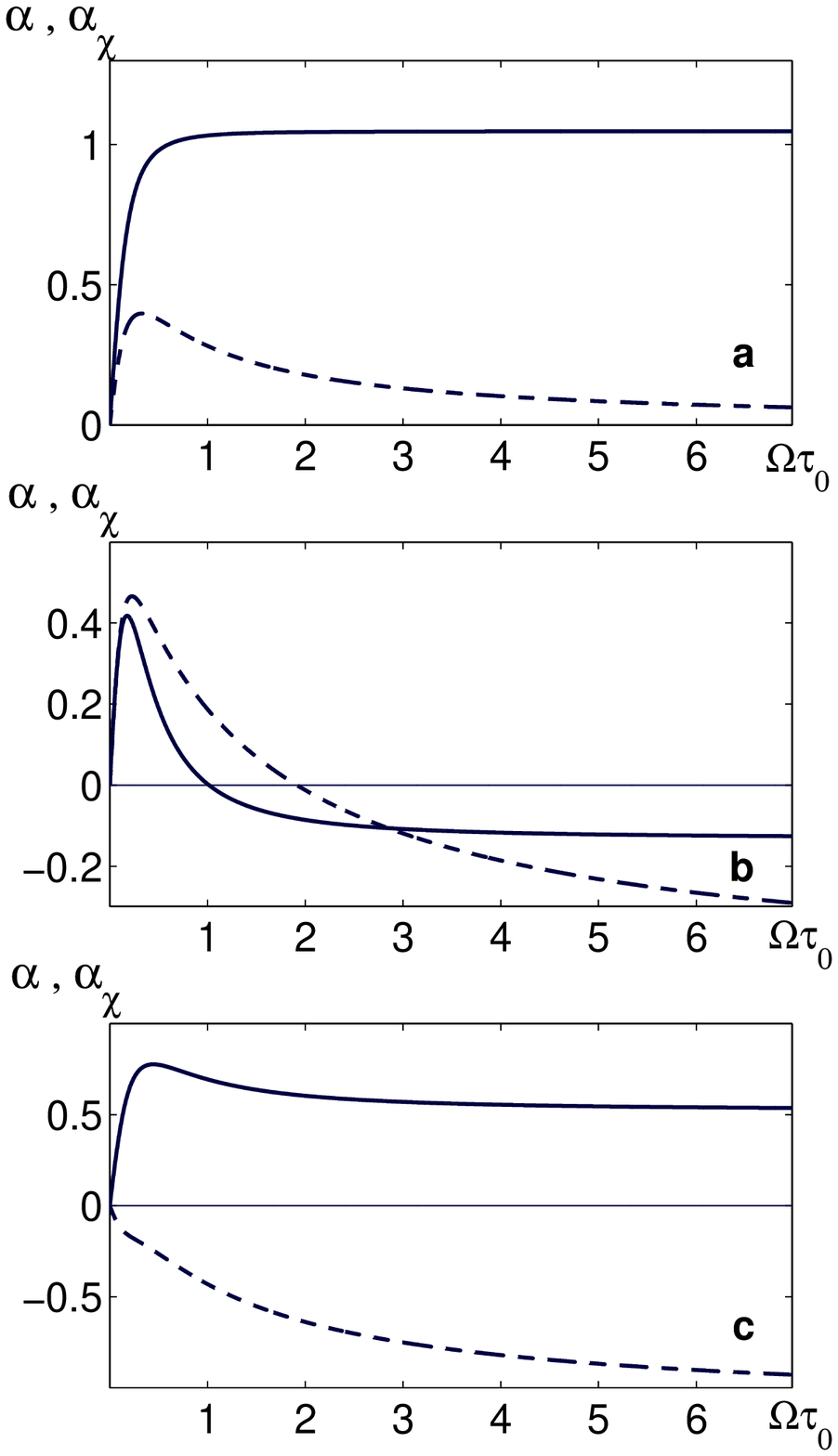}
\caption{\label{Fig4} The $\alpha$ (solid line) and $
\alpha_{\chi} = - (1/3) \tau_{_{0}} \chi^{(v)} $ (dashed line) as
the functions of the parameter $ \Omega \tau_{_{0}} $ for $
\phi_{l} = 90^{\circ} $ and different values of the degrees of
anisotropy: (a). $\varepsilon = 0$ and $\sigma = 1$; (b).
$\varepsilon = 1.2 $ and $\sigma = 2$; (c). $\varepsilon = 13$ and
$\sigma = 0 .$ In FIG. 4b the $\alpha$ effect is multiplied by $ 2
$ and in FIG. 4c by $5/2 .$}
\end{figure}

Figure 1b shows the ranges of parameters $ (\sigma $ and $
\lambda) $ with different behavior of $ \alpha_{\chi} .$ In
FIG.~1b these ranges are separated by lines $ \sigma = (9/41)(25
\lambda / 12 - 1) $ and $ \lambda = 0 ,$ where $ - 2 < \lambda < 2
$ and $ - 9/2 < \sigma < 3 .$ The numeration of the ranges in
FIG.~1b for $ \alpha_{\chi} $ is the same as for the parameter
$\alpha$ in FIG. 1a. Comparison of FIGS. 1a and 1b shows that the
ranges III and IV (whereby the $\alpha$-effect changes its sign at
a certain value of $ \Omega \tau_{_{0}} $ and a certain range of
the latitudes $ \phi_l) $ do not exist for $ \alpha_{\chi} .$ On
the other hand, there is a new range (the range VI) in FIG 1b
whereby the sign of $ \alpha_{\chi} $ changes from negative value
for a slow rotation to positive value for $ \omega \gg 1 .$ The
locations of the ranges II and V for $ \alpha_{\chi} $ are
different from that of the $\alpha$-effect. Therefore, the
behavior of the parameter $ \alpha_{\chi} $ and the
$\alpha$-effect are different in a rotating fluid.

The dependencies of the $\alpha$ effect on the latitude $ \phi_{l}
$ for different values of the degrees of anisotropy $ \varepsilon
$ and $\sigma$, and different values of the parameter $ \Omega
\tau_{_{0}} $ are shown in FIG. 5. It is seen in FIG. 5b that the
$\alpha$ effect changes its sign at $ \phi_{l} \approx 20^{\circ}
- 40^{\circ} $ for $ \Omega \tau_{_{0}} = 5 $ (this value of $
\Omega \tau_{_{0}} $ corresponds to the lower part of the solar
convective zone).

\begin{figure}
\centering
\includegraphics[width=8cm]{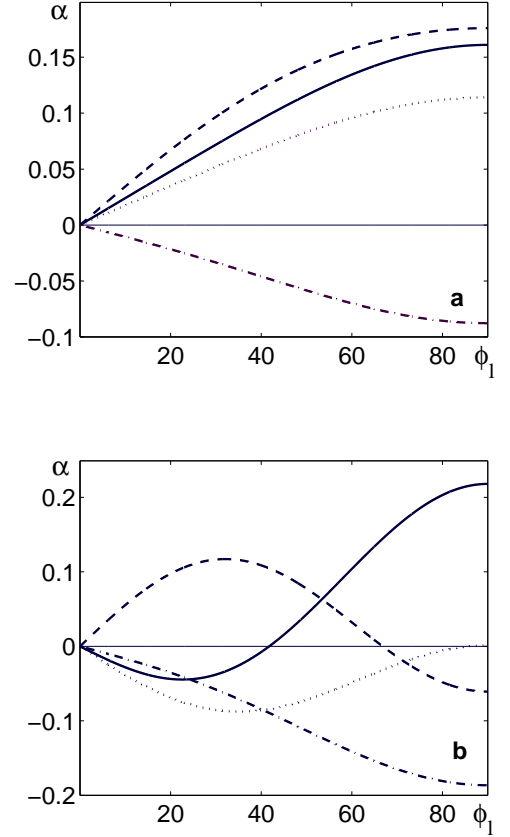}
\caption{\label{Fig5} The dependence of the $\alpha$ effect on the
latitude $ \phi_{l} $ for different values of the degrees of
anisotropy: $\varepsilon = 1.2$ and $\sigma = 2 $ (dashed);
$\varepsilon = 13 $  and $\sigma = 0$ (solid); $\varepsilon = 13$
and $\sigma = 2.2 $ (dashed-dotted); $\varepsilon = 13$ and
$\sigma = 0.415 $ (dotted); and for different values of the
parameter $ \Omega \tau_{_{0}} :$ (a). $ \Omega \tau_{_{0}} =
0.1$; (b). $ \Omega \tau_{_{0}} = 5 .$ The dashed-dotted line in
FIG. 5b shows $\alpha / 5 $. The latitude is measured in degrees.}
\end{figure}

In view of the applications to the astrophysics, the case with
negative $\alpha$-effect for $ \phi_{l} > 0 $ is most important
because this provides a propagation of the solar dynamo waves to
the equator according to the solar observations (see, {\em e.g.,}
\cite{M78,P79,KR80,ZRS83,S89}).

\subsection{The effective drift velocity of the mean magnetic field}

Now we determine the effective drift velocity $ V_{k}^{(d)} \equiv
- (1/2) \varepsilon_{kij} a_{ij}^{(AS)} = V_{k}^{(1)} +
V_{k}^{(2)} $ of the mean magnetic field using Eq.~(\ref{E1}),
where $ a_{ij}^{(AS)} = (1/2) (a_{ij} - a_{ji}) ,$ and
\begin{eqnarray}
{\bf V}^{(1)} &=& {1 \over 48} \biggl({l_{0} u_{0} \over L_{\rho}}
\biggr) \{{\bf e} [E_{1}(\omega) + E_{2}(\omega) \sin^2 \phi_l
\nonumber \\
& & + E_{3}(\omega) \sin^4 \phi_l] - {1 \over 2} {\bf e}_{\theta}
[E_{4}(\omega)
\nonumber \\
& & + E_{3}(\omega) \sin^2 \phi_l] \sin (2 \phi_l) \} \;,
\label{E4} \\
{\bf V}^{(2)} &=& {1 \over 6} \biggl({l_{0}^{2} \Omega \over
L_{\rho}} \biggr) [E_{5}(\omega)
\nonumber \\
& & + E_{6}(\omega) \sin^2 \phi_l] (\bec{\hat \omega} {\bf \times}
{\bf e}) \; . \label{E15}
\end{eqnarray}
(for details, see Appendix A), where $ r, \theta, \varphi $ are
the spherical coordinates, $ \phi_l = \pi / 2 - \theta ,$ $ \,
\bec{\hat \omega} {\bf \times} {\bf e} = \cos \phi_l \, {\bf
e}_{\varphi} ,$ $ \bec{\hat \omega} = {\bf e} \, \sin \phi_l -
{\bf e}_{\theta} \, \cos \phi_l ,$ the functions $ E_{k}(\omega) $
are given in Appendix C. For a slow rotation $ (\omega \ll 1) $
the effective drift velocities are given by
\begin{eqnarray}
{\bf V}^{(1)} &\approx& - {4 \over 15} \biggl({l_{0} u_{0} \over
L_{\rho}} \biggr) \biggl[{\bf e} \biggl(1 - {\sigma \over 6} + {5
\over 4(\varepsilon + 2)}
\nonumber\\
& & + O(\omega^2) \biggr)  - {\bf e}_{\theta} {5 \over 12}
\omega^2 \biggl(1 - {\sigma \over 6}
\nonumber\\
& & - {3 \varepsilon - 1 \over 7(\varepsilon + 2)} \biggr) \sin (2
\phi_l) \,  \biggr] \;,
\label{E5}\\
{\bf V}^{(2)} &\approx& {4 \over 5} \biggl({l_{0}^{2} \Omega \over
L_{\rho}} \biggr) \biggl(1 - {\sigma \over 6} + {5(1 -
\varepsilon) \over 9(\varepsilon + 2)} \biggr) (\bec{\hat \omega}
{\bf \times} {\bf e}) \;,
\label{E16}
\end{eqnarray}
and for $ \omega \gg 1 $ they are given by
\begin{eqnarray}
{\bf V}^{(1)} &\approx& - {\pi \over 8 \omega} \biggl({l_{0} u_{0}
\over L_{\rho}} \biggr) [{\bf e}(1 +  \sin^2 \phi_l)
\nonumber\\
& & - {1 \over 2} {\bf e}_{\theta} \, \sin (2 \phi_l)] \;,
\label{E6} \\
{\bf V}^{(2)} &\approx& {1 \over 2 \omega} \biggl({l_{0} u_{0}
\over L_{\rho}} \biggr) (\sigma - 1) \sin^{2} \phi_l \, \,
(\bec{\hat \omega} {\bf \times} {\bf e}) \; . \label{E17}
\end{eqnarray}
For $ \omega = 0 $ this effective drift velocity, $ {\bf V}^{(1)}
,$ corresponds to the well-known turbulent diamagnetic velocity
(see, {\em e.g.,} \cite{M78,P79,KR80,ZRS83,RSS88}). Indeed, since
we suggested that $ \bec{\nabla} (\rho \langle {\bf u}^{2}
\rangle) \approx 0 ,$ thus $ \bec{\nabla} \langle {\bf u}^{2}
\rangle / \langle {\bf u}^{2} \rangle \approx L_{\rho}^{-1} {\bf
e} $ and Eq.~(\ref{E5}) for $ \omega = 0 $ reads
\begin{eqnarray}
{\bf V}^{(1)} \approx - {4 \over 15} \biggl({l_{0}^{2} \Omega
\over L_{\rho}} \biggr) \biggl(1 - {\sigma \over 6} + {5 \over
4(\varepsilon + 2)} \biggr) \, \tau_{_{0}} \, \bec{\nabla} \langle
{\bf u}^{2} \rangle \;, \label{E5E}
\end{eqnarray}
where $ \tau_{_{0}} = l_{0} / u_{0} .$ Figure 6 shows the
effective drift velocities: $V_\theta^{(1)}$ and $V_r^{(1)}$ as
the functions of the parameter $ \Omega \tau_{_{0}} $ for
different values of the degrees of anisotropy.

\begin{figure}
\centering
\includegraphics[width=8cm]{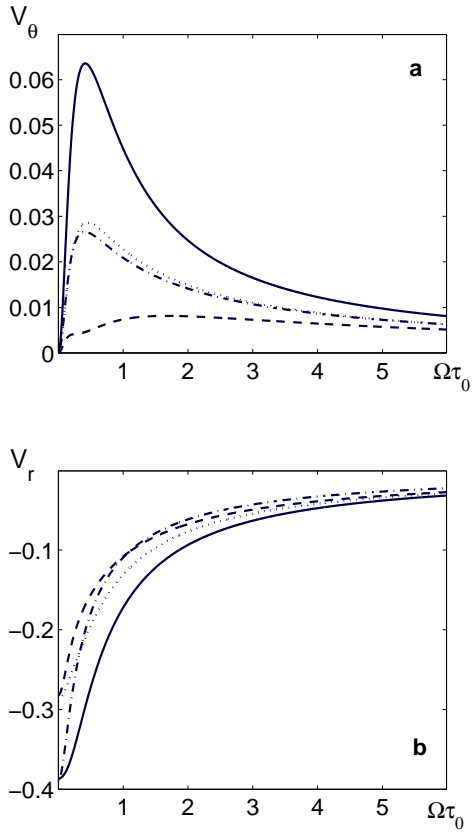}
\caption{\label{Fig6} The effective drift velocities: (a).
$V_\theta^{(1)}$ for $ \phi_{l} = 45^{\circ} $; (b). $V_r^{(1)}$
for $ \phi_{l} = 90^{\circ} $ as the functions of the parameter $
\Omega \tau_{_{0}} $ for different values of the degrees of
anisotropy: $\varepsilon = 13$ and $\sigma = -2.2 $ (solid);
$\varepsilon = 1.2 $ and $\sigma = 2$ (dashed); $\varepsilon = 13$
and $\sigma = 0 $ (dotted); $\varepsilon = 0$ and $\sigma = 1$
(dashed).}
\end{figure}

The effective drift velocity $ {\bf V}^{(2)} $ causes an
additional differential rotation. Indeed, let us introduce the
angular velocity difference, $ \delta \Omega ,$ which is
determined from the identity: $ {\bf V}^{(2)} = \delta \Omega \, r
\, (\bec{\hat \omega} {\bf \times} {\bf e}) .$ Comparison of this
definition with Eqs. (\ref{E16}) and (\ref{E17}) yields equations
for $ \delta \Omega (r) \propto r^{-1} .$ Calculating the $ r
$-derivatives of $ \delta \Omega (r) $ we obtain equations which
determine the differential rotation for a slow rotation $ (\omega
\ll 1) $
\begin{eqnarray}
{\partial (\delta \Omega) \over \partial r} \approx - {4 \over 5}
\biggl({l_{0}^{2} \Omega \over L_{\rho} \, r^{2}} \biggr) \biggl(1
- {\sigma \over 6} + {5(1 - \varepsilon) \over 9(\varepsilon + 2)}
\biggr) \;,
\label{E18}
\end{eqnarray}
and for $ \omega \gg 1 $:
\begin{eqnarray}
{\partial (\delta \Omega) \over \partial r} \approx {1 \over 2
\omega} \biggl({l_{0} u_{0} \over L_{\rho} \, r^{2}} \biggr) (1 -
\sigma) \sin^{2} \phi_l \; .
\label{E19}
\end{eqnarray}

The electromotive force has a term $ a_{ij}^{(c)} B_{j} $ which
for an axisymmetric case contributes only to an additional effective drift
velocity, $ {\bf V}^{(3)} ,$ of the mean magnetic field, {\em i.e.,}
\begin{eqnarray}
a_{ij}^{(c)} B_{j} = [{\bf V}^{(3)} {\bf \times} ({\bf B}_{p} -
{\bf B}_{T})]_{i} \;
\label{E7}
\end{eqnarray}
(for details, see Appendix A), where $ {\bf B} = {\bf B}_{T} +
{\bf B}_{p} $ is the mean magnetic field with the toroidal, $ {\bf
B}_{T}, $ and the poloidal, $ {\bf B}_{p} ,$ components, the
tensor $ a_{ij}^{(c)} $ is determined by Eq.~(\ref{FF9}), and the
additional effective drift velocity is given by
\begin{eqnarray}
{\bf V}^{(3)}  &=& {1 \over 24} \biggl({l_{0} u_{0} \over
L_{\rho}} \biggr) \{ {\bf e} [E_{7}(\omega) + E_{8}(\omega) \sin^2
\phi_l
\nonumber \\
& & + E_{9}(\omega) \sin^4 \phi_l] + {1 \over 2} {\bf e}_{\theta}
[E_{10}(\omega)
\nonumber \\
& & - E_{9}(\omega) \sin^2 \phi_l] \sin (2 \phi_l) \} \label{E8}
\end{eqnarray}
(for details, see Appendix A). Note that for a slow rotation $
(\omega \ll 1) $ the additional effective drift velocity is very
small, {\em i.e.,} $ V^{(3)} \sim O(\omega^{2}) ,$ and for $
\omega \gg 1 $ it is given by
\begin{eqnarray}
{\bf V}^{(3)} & \approx & {\pi \over 8 \omega} \biggl({l_{0} u_{0}
\over L_{\rho}} \biggr) \cos \phi_l \, \{ {\bf e} \, \cos \phi_l
[\lambda + 10 - 13 \sigma
\nonumber \\
& & - 18 (\sigma - 1) \sin^2 \phi_l] + {\bf e}_{\theta} \, \sin
\phi_l [\lambda - \sigma
\nonumber \\
& & - 18 (\sigma - 1) \cos^2 \phi_l] \} \; .
\label{E11}
\end{eqnarray}
Now we determine the total effective drift velocity in an
axisymmetric case:
\begin{eqnarray}
[{\bf V}^{(d)} {\bf \times} {\bf B}]_{i} + a_{ij}^{(c)} B_{j}
= [{\bf V}^{(B)} {\bf \times} {\bf B}_{T} + {\bf V}^{(A)} {\bf \times}
{\bf B}_{p}]_{i} \;,
\label{E9}
\end{eqnarray}
where
\begin{eqnarray}
{\bf V}^{(B)} = {\bf V}^{(d)} - {\bf V}^{(3)} \;, \quad
{\bf V}^{(A)} = {\bf V}^{(d)} + {\bf V}^{(3)}  \; .
\label{E10}
\end{eqnarray}
Therefore, the effective drift velocities, $ {\bf V}^{(B)} $ and $
{\bf V}^{(A)} ,$ for the toroidal and poloidal magnetic fields are
different. The additional effective drift velocity, $ {\bf
V}^{(3)} ,$ is a result of an interaction of turbulent convection
with inertial waves and Rossby waves. Indeed, a part of the tensor
$ a_{ij}^{(c)}({\bf k}) \propto \psi_{_{\Omega}} \psi_{_{R}} ,$
where $ \psi_{_{\Omega}} = 2 ({\bf \Omega} \cdot {\bf k}) / k $ is
the frequency of the inertial waves and $ \psi_{_{R}} = 2 \Lambda
\Omega_{x} k_{y} / k^{2} $ is the frequency of Rossby waves [see
Eqs. (\ref{E7}) and (\ref{D5})].

\section{Discussion}

In this paper we studied an effect of rotation on a developed
turbulent stratified convection. This allowed us to determine the
dependencies of the hydrodynamic helicity, the alpha-tensor and
the effective drift velocity of the mean magnetic field on the
rate of rotation and an anisotropy of turbulence. We demonstrated
that in a turbulent convection the alpha-effect can change its
sign depending on the rate of rotation and an anisotropy of
turbulence. We found different properties of the effective drift
velocity of the mean magnetic field in a rotating turbulent
convection. In particular, a poloidal effective drift velocity can
be diamagnetic or paramagnetic depending on the rate of rotation.
There is a difference in the effective drift velocities for the
toroidal and poloidal magnetic fields which increases with the
rate of rotation. We found also a toroidal effective drift
velocity which can play a role of an additional differential
rotation.

Some of the results obtained in our paper using the
$\tau$-approximation are observed in the direct numerical
simulations of the stratified turbulent convection (see
\cite{OSB02}).  In particular, it was found in \cite{OSB02} that
the alpha-effect can change its sign depending on the rate of
rotation. It was also demonstrated in \cite{OSB02} that there is a
difference in the effective drift velocities for the toroidal and
poloidal magnetic fields, and that an observed toroidal effective
drift velocity in \cite{OSB02} can play a role of an additional
differential rotation.

Now we apply the obtained results for the analysis of an axisymmetric
$\alpha \Omega$-dynamo. The mean magnetic field in an axisymmetric case is
given by $ {\bf B} = B {\bf e}_{\varphi} + \bec{\nabla} {\bf \times}
(A {\bf e}_{\varphi}) ,$ where $ A $ is the vector potential.
The equations for $ B $ and $ A $ in dimensionless form are given by
\begin{eqnarray}
{\partial B \over \partial t} + r_{\perp} \, \bec{\nabla} {\bf
\cdot} ( {\bf V}^{(B)} \, r_{\perp}^{-1} \, B) &=& D \, (\hat
\Omega A) + \Delta_{s} B \;,
\label{L10} \\
{\partial A \over \partial t} + r_{\perp}^{-1} \, ({\bf V}^{(A)}
{\bf \cdot} \bec{\nabla}) \, (r_{\perp} \, A) &=& \alpha B +
\Delta_{s} A \;, \label{L11}
\end{eqnarray}
where the length is measured in units of the thickness of the
convective zone $ L_{c} ,$ the time is measured in units of $
L_{c}^{2} / \eta_{_{T}} ,$ the velocity is measured in units of $
\eta_{_{T}} / L_{c} ,$ the turbulent magnetic diffusion $
\eta_{_{T}} = l_{0} u_{0} / 3,$ and $u_{0}$ is the characteristic
turbulent velocity in the scale $ l_{0} ,$ $ \,  D = R_\alpha
R_\omega $ is the dynamo number, $ R_\alpha = L_{c} \alpha_\ast /
\eta_{_{T}} $ and $ R_\omega = L_{c}^{2} (\delta \Omega)_\ast /
\eta_{_{T}} .$ Here $ \alpha $ is measured in units of the maximum
value $ \alpha_\ast $ of the $ \alpha $ effect, $ (\delta
\Omega)_\ast $ is the characteristic differential rotation in the
scale $ L_{c} ,$ $ \hat \Omega A = [\bec{\nabla} (\delta \Omega)
{\bf \times} \bec{\nabla} (r_{\perp} \, A)] \cdot {\bf
e}_{\varphi} ,$ $ \, \Delta_{s} = \Delta - r_{\perp}^{-2} ,$ $
r_{\perp} = r \sin \theta $ and we used the induction equation for
the mean magnetic field (see, {\em e.g.,}
\cite{M78,P79,KR80,ZRS83,RSS88}) and Eqs. (\ref{E9}) and
(\ref{E10}). When $ {\bf V}^{(A)} = {\bf V}^{(B)} $ and $
\bec{\nabla} {\bf \cdot} {\bf V}^{(B)} = 0 ,$ Eqs. (\ref{L10}) and
(\ref{L11}) coincide with that given in \cite{M78}. Now we seek
for a solution of Eqs. (\ref{L10}) and (\ref{L11}) in the form $
A, B \propto \exp(\hat \gamma t + i {\bf k} {\bf \cdot} {\bf x})
,$ where $ {\bf k} = k {\bf e}_{k} ,$ $ \, {\bf e}_{k} = {\bf
e}_{\varphi} {\bf \times} {\bf e}_{_{\Omega}} ,$ the unit vector $
{\bf e}_{_{\Omega}} $ is directed opposite to $ \bec{\nabla}
(\delta \Omega) $ and
\begin{eqnarray}
\hat \gamma &=& \kappa / 2 - k^{2} - i k U^{(1)}
\nonumber \\
& & \pm [(\kappa / 2 + i k U^{(3)})^{2} + i k D]^{1/2} \;,
\label{L12}
\end{eqnarray}
$ \kappa = - \bec{\nabla} {\bf \cdot} {\bf V}^{(B)} ,$
$ \, U^{(1,3)} = {\bf V}^{(1,3)} \, {\bf \cdot} \, {\bf e}_{k} $
and $ \hat \gamma = \gamma_{_{B}} + i \omega_{_{B}} .$ In the limit of large
dynamo number $ | D | $ the maximum growth rate of the mean magnetic
field $ \gamma_{_{B}} $ is given by
\begin{eqnarray}
\gamma_{_{B}} = (3/4) (| D | / 4)^{2/3} + \kappa / 2 \;,
\label{L14}
\end{eqnarray}
which is achieved at the wave number $ k_{m} = (1/2) (| D | / 4)^{1/3} .$
At this wave number the frequency $ \omega_{_{B}} $ of the dynamo wave is
\begin{eqnarray}
\omega_{_{B}} = - (| D | / 4)^{2/3}  - (1/2) U^{(1)} (| D | /
4)^{1/3} \; \label{L15}
\end{eqnarray}
(see \cite{KRS01}). The negative sign of $ \omega_{_{B}} $ implies
that the dynamo waves propagate to the equator in agreement with
the solar magnetic field observations. On the other hand, the
divergence of the effective drift velocity $ {\bf V}^{(B)} $ of
the toroidal magnetic field can cause an increase of the growth
rate of the mean magnetic field when $ \kappa > 0 .$ The change of
the sign of the $\alpha$-effect depending on the rate of rotation
and anisotropy of turbulent convection (see Section III-B) can
explain the observed direction of propagation of the solar dynamo
waves.

Note that a meridional circulation in the solar convective zone
can also cause an equatorward drift of the solar dynamo wave (see,
e.g., \cite{M78,DC94,CSD95}). However, it was shown recently in
\cite{DBE02} that the meridional velocity, which is required for
the equatorward propagation of the solar dynamo wave with the
period $ \sim 22 $ years, should be of the order of $ \sim 10-12 $
m/s. Such large meridional velocities are not observed on the
solar surface. On the other hand, we found that the effective
drift velocities of the mean magnetic field have a meridional
component (along $ {\bf e}_\theta )$. This velocity has the
maximum $ (V_\theta^{(1)})_{\rm max} \sim 10-12 $ m/s in the upper
part of the solar convective zone. Therefore, this meridional
effective drift velocity of the mean magnetic field can cause the
equatorward propagation of the solar dynamo wave in the upper part
of the solar convective zone. Note that the meridional
circulations in the solar convection zone and the meridional
component of the effective drift velocities of the mean magnetic
field are different characteristics, because the first velocity
describes large-scale fluid motions (which may cause advection of
the mean magnetic field by the large-scale fluid motions, i.e., by
the mean flow), and the second velocity determines the drift
velocity of the mean magnetic field (which is originated from the
mean electromotive force $ \bec{\cal E} = \langle {\bf u} \times
{\bf b} \rangle ).$

We found also that in the upper part of the solar convective zone
the $\alpha$ effect does not change its sign, i.e., it is
positive. But in the lower part of the solar convective zone the
$\alpha$ effect changes its sign, because the parameter $\Omega
\tau_{_{0}}$ increases with the increase of the depth the solar
convective zone, and the $\alpha$ effect becomes negative.
Therefore, in the lower part of the solar convective zone the
negative $\alpha$ effect is responsible for the equatorward
propagation of the solar dynamo waves. On the other hand, the
meridional effective drift velocity of the mean magnetic field in
the lower part of the solar convective zone is very small and,
thus, it cannot be used for the explanation of the equatorward
propagation of the solar dynamo wave.

Therefore, both effects, the meridional effective drift velocity
of the mean magnetic field in the upper part of the solar
convective zone and the sign reversal of the $\alpha$ effect in
the lower part of the solar convective zone, can cause the
equatorward propagation of the solar dynamo wave.

Note that in the present study we did not discuss the magnetic
buoyancy effects which play an important role in a creation of
strongly inhomogeneous magnetic structures (see, e.g.,
\cite{P79,KMR96,KR94,MFS92,FSS94}).

\begin{acknowledgments}
We have benefited from numerous discussions on the effect of
rotation on turbulence with K.-H. R\"{a}dler. This work was
partially supported by INTAS Program Foundation (Grant No.
99-348).
\end{acknowledgments}

\appendix

\section{Derivations of Eqs. (\ref{D10}), (\ref{F12}),
(\ref{E4}), (\ref{E15}) and (\ref{E8}).}

\subsection{The conservation equations}

Equations~(\ref{B1}) and (\ref{B2}) yield the following
conservation equations for the kinetic energy $ W_{u} =
\rho_{_{0}} {\bf u}^{2} / 2 $ and for $ W_{S} = \rho_{_{0}} S^{2}
/ 2 :$
\begin{eqnarray}
\partial  W_{u} / \partial t + \bec{\nabla} \cdot {\bf F}_{u}
&=& I_{u} - D_{u} \;,
\label{P2} \\
\partial  W_{S} / \partial t + \bec{\nabla} \cdot {\bf F}_{S}
&=& I_{S} - D_{S} \;, \label{P3}
\end{eqnarray}
where the source terms in these equations are $ I_{u} = -
\rho_{_{0}} ({\bf u} \cdot {\bf g}) S $ and $ I_{S} = - I_{u}
\tilde \Omega_{b}^{2} / g^{2} ,$ the dissipative terms are $ D_{u}
= - \rho_{_{0}} ({\bf u} \cdot {\bf f}_{\nu}) $ and $ \, D_{S} =
\rho_{_{0}} S \, \bec{\nabla} \cdot {\bf F}_{\kappa} ,$ the fluxes
are $ {\bf F}_{u} = {\bf u} \, (W_{u} + P) $ and $ {\bf F}_{S} =
{\bf u} W_{S} .$ Equations~(\ref{P2}) and (\ref{P3}) yield a
conservation equation for $ W_{E} = W_{u} \tilde\Omega_{b}^{2} /
g^{2} + W_{S}$
\begin{eqnarray}
\partial  W_{E} / \partial t + \bec{\nabla} \cdot {\bf F}_{E}
&=& - D_{E} \;, \label{P4}
\end{eqnarray}
where the dissipative term is $ D_{E} = D_{u} \tilde\Omega_{b}^{2}
/ g^{2}  + D_{S} $ and the flux is $ {\bf F}_{E} = {\bf F}_{u}
\tilde\Omega_{b}^{2} / g^{2} + {\bf F}_{S} .$ Equation~(\ref{P4})
does not have a source term and it implies that without the
dissipation $ (D_{E} = 0) $ the value $ \int W_{E} \, dV $ is
conserved, where in the latter formula the integration over the
volume is performed. For the convection $ \tilde\Omega_{b}^{2} < 0
$ and, therefore, $ W_{S} \approx W_{u} |\tilde\Omega_{b}^{2}| /
g^{2} .$ Averaging Eq.~(\ref{P2}) over an ensemble of fluctuations
we obtain a relationship between the flux of the entropy and the
dissipation of the kinetic energy in a stationary turbulent
convection: $ \langle u_{i} S \rangle g_{i} = \langle {\bf u}
\cdot {\bf f}_{\nu} \rangle .$ Similarly, averaging Eq.~(\ref{P4})
over an ensemble of fluctuations we obtain $ \langle {\bf u}^{2}
\rangle = \langle S^{2} \rangle    (g^{2} /
|\tilde\Omega_{b}^{2}|) .$ Equation~(\ref{P2}) yields the
relationship between $ \Phi^{\ast}_z $ and $ f_{\ast} $: $ \,
f_{\ast} = 2 \lambda g \tau_{_{0}} \Phi^{\ast}_z / \varepsilon .$

\subsection{Modification of turbulent convection by rotation}

Now we study a modification of turbulent convection by rotation.
To this end we derive equations for the following second moments:
\begin{eqnarray*}
f_{ij}({\bf k}) &=& \hat L(v_{i},v_{j}) \;, \quad \chi({\bf k}) =
\hat L(w,v_{z}) \;,
\\
F({\bf k}) &=& \hat L(s,w) \;, \quad G({\bf k}) = \hat L(w,w) \;,
\\
\Phi_i({\bf k}) &=& \hat L(s,v_{i}) \;, \quad \Theta({\bf k}) =
\hat L(s,s) \;,
\end{eqnarray*}
using Eqs. (\ref{B4})-(\ref{B6}), where $ \hat L(a,b) = \langle
a({\bf k}) b(-{\bf k}) \rangle $ and $ {\bf v} = \sqrt{\rho_0(z)}
\, {\bf u} .$ The equations for these correlation functions are
given by
\begin{eqnarray}
{\partial f({\bf k}) \over \partial t} &=& {2 \psi_{_{\Lambda}}
\over k^{2}} \chi_{_{R}}({\bf k}) - {2 \psi_{_{\Omega}} \over k}
\chi_{_{I}}({\bf k})
\nonumber\\
& & + 2 g_{\perp}({\bf k}) \Phi_{R}({\bf k}) + f_{_{N}} \;,
\label{B13} \\
{\partial \chi({\bf k}) \over \partial t} &=& (ik \psi_{_{\Omega}}
- \psi_{_{\Lambda}}) \biggl(f({\bf k}) - {1 \over k^{2}} G({\bf
k})\biggr)
\nonumber\\
& & - i \psi_{_{R}} \chi({\bf k}) + g_{\perp}({\bf k}) F(-{\bf k})
+ \chi_{_{N}} \;,
\label{B14} \\
{\partial \Phi_z({\bf k}) \over \partial t} &=& - {\Omega_{b}^{2}
\over g} f({\bf k}) - {1 \over k^{2}}(i k \psi_{_{\Omega}} -
\psi_{_{\Lambda}}) F({\bf k})
\nonumber\\
& & - i \psi_{_{R}} \Phi_z({\bf k}) + g_{\perp}({\bf k})
\Theta({\bf k}) + \Phi_{N} \;,
\label{B15} \\
{\partial F({\bf k}) \over \partial t} &=& - {\Omega_{b}^{2} \over
g} \chi(-{\bf k}) - (i k \psi_{_{\Omega}} + \psi_{_{\Lambda}})
\Phi_z({\bf k})
\nonumber\\
& & + F_{N} \;,
\label{B16} \\
{\partial G({\bf k}) \over \partial t} &=& 2k \psi_{_{\Omega}}
\chi_{_{I}}({\bf k}) - 2 \psi_{_{\Lambda}} \chi_{_{R}}({\bf k}) +
G_{N} \;,
\label{B17} \\
{\partial \Theta({\bf k}) \over \partial t} &=& - 2
{\Omega_{b}^{2} \over g} \Phi_{R}({\bf k}) + \Theta_{N} \;,
\label{B18}
\end{eqnarray}
where
\begin{eqnarray*}
\Phi_{R}({\bf k}) &=& [\Phi_z({\bf k}) + \Phi_z(-{\bf k})] / 2 \;,
\\
\Phi_{I}({\bf k}) &=& [\Phi_z({\bf k}) - \Phi_z(-{\bf k})] / 2 i
\;,
\end{eqnarray*}
and similarly for other second moments, $ f_{_{N}} ,$ $
\chi_{_{N}} , \ldots ,$ $ \Theta_{N} $ are the third moments which
are given by
\begin{eqnarray*}
f_{_{N}}({\bf k}) &=& \hat L(V_{N},v_{z}) + \hat L(v_{z},V_{N})
\;,
\\
\chi_{_{N}}({\bf k}) &=& \hat L(W_{N},v_{z}) + \hat L(w,V_{N}) \;,
\\
\Phi_{N}({\bf k}) &=& \hat L(S_{N},v_{z}) + \hat L(s,V_{N}) \;,
\\
F_{N}({\bf k}) &=& \hat L(S_{N},w) + \hat L(s,W_{N}) \;,
\\
G_{N}({\bf k}) &=& \hat L(W_{N},w) + \hat L(w,W_{N}) \;,
\\
\Theta_{N}({\bf k}) &=& \hat L(S_{N},s) + \hat L(s,S_{N}) \;,
\end{eqnarray*}
and
\begin{eqnarray}
V_{N} &=& - \sqrt{\rho_0} \, {\bf e} \cdot \{\bec{\nabla} {\bf
\times} [\bec{\nabla} {\bf \times} (({\bf u} \cdot \bec{\nabla})
{\bf u} - {\bf f}_\nu)] \},
\label{RD1} \\
W_{N} &=& \sqrt{\rho_0} \, {\bf e} \cdot [\bec{\nabla} {\bf
\times} ({\bf u} {\bf \times} {\bf w} + {\bf f}_\nu)] \;,
\label{RD2} \\
S_{N} &=& - \sqrt{\rho_0} \, \biggl\{({\bf u} \cdot \bec{\nabla})
\biggl({s \over \sqrt{\rho_0}}\biggr)
\nonumber\\
& & + {1 \over T_{0}} \, {\rm div} \, \biggl[{\bf
F}_{\kappa}\biggl({s \over \sqrt{\rho_0}}\biggr) \biggr] \biggr\}
\;, \label{RD3}
\end{eqnarray}
$ \psi_{_{\Lambda}} = {\bf \Omega} \cdot {\bf \Lambda} ,$ $ \quad
\psi_{_{\Omega}} = 2  ({\bf \Omega} \cdot {\bf k}) / k ,$ $ \quad
\psi_{_{R}} = 2 \Lambda \Omega_{x} k_{y} / k^{2} ,$ and $
g_{\perp}({\bf k}) = g (k_{\perp} / k)^{2} .$ We assumed that $
(1/4) \Lambda^{2} \ll k^{2} .$ Now we introduce the following
variables:
\begin{eqnarray*}
\chi_{p}({\bf k}) &=& k \psi_{_{\Omega}} \, \chi_{_{R}}({\bf k}) +
\psi_{_{\Lambda}} \, \chi_{_{I}}({\bf k}) \;,
\\
\chi_{m}({\bf k}) &=& k \psi_{_{\Omega}} \, \chi_{_{I}}({\bf k}) -
\psi_{_{\Lambda}} \, \chi_{_{R}}({\bf k}) \;,
\\
F_{p}({\bf k}) &=& k \psi_{_{\Omega}} \, F_{R}({\bf k}) -
\psi_{_{\Lambda}} \, F_{I}({\bf k}) \;,
\\
F_{m}({\bf k}) &=& k \psi_{_{\Omega}} \, F_{I}({\bf k}) +
\psi_{_{\Lambda}} \, F_{R}({\bf k}) \;,
\end{eqnarray*}
which allow us to rewrite Eqs. (\ref{B13})-(\ref{B18}) as follows
\begin{eqnarray}
{\partial f({\bf k}) \over \partial t} &=& - {2 \over k^{2}}
\chi_{m}({\bf k}) + 2 g_{\perp}({\bf k}) \Phi_{R}({\bf k})
\nonumber\\
& & + f_{_{N}} \;,
\label{EB13} \\
{\partial \chi_{p}({\bf k}) \over \partial t} &=& \psi_{_{R}}
\chi_{m}({\bf k}) + g_{\perp}({\bf k}) F_{p}({\bf k}) +
\chi_{_{N}}^{(p)} \;,
\label{EB14} \\
{\partial \chi_{m}({\bf k}) \over \partial t} &=& (k
\psi_{_{\Omega}})^2 \biggl(f({\bf k}) - {1 \over k^{2}} G({\bf
k})\biggr) - \psi_{_{R}} \chi_{p}({\bf k})
\nonumber\\
& & - g_{\perp}({\bf k}) F_{m}({\bf k}) + \chi_{_{N}}^{(m)} \;,
\label{EB12} \\
{\partial \Phi_{R}({\bf k}) \over \partial t} &=& {1 \over k^{2}}
F_{m}({\bf k}) + \psi_{_{R}} \Phi_{I}({\bf k})
\nonumber\\
& & + g_{\perp}({\bf k}) \Theta({\bf k}) + \Phi_{N}^{(R)} \;,
\label{EB15} \\
{\partial \Phi_{I}({\bf k}) \over \partial t} &=& - {1 \over
k^{2}} F_{p}({\bf k}) - \psi_{_{R}} \Phi_{R}({\bf k}) +
\Phi_{N}^{(I)} \;,
\label{EB11} \\
{\partial F_{p}({\bf k}) \over \partial t} &=&  (k
\psi_{_{\Omega}})^2 \Phi_{I}({\bf k}) + F_{N}^{(p)} \;,
\label{EB16} \\
{\partial F_{m}({\bf k}) \over \partial t} &=& - (k
\psi_{_{\Omega}})^2 \Phi_{R}({\bf k}) + F_{N}^{(m)} \;,
\label{EB10} \\
{\partial G({\bf k}) \over \partial t} &=& {2 \over k^{2}}
\chi_{m}({\bf k}) + G_{N} \;,
\label{EB17} \\
{\partial \Theta({\bf k}) \over \partial t} &=& \Theta_{N} \;,
\label{EB18}
\end{eqnarray}
where we neglected small terms proportional to $ \Omega_{b}^{2} /
g .$

Next, we use the $ \tau $-approximation which allows us to express
the third moments $ f_{_{N}} ,$ $ \chi_{_{N}}^{(p)}, \ldots ,$ $
\Theta_{N} $ in Eqs. (\ref{EB13})-(\ref{EB18}) in terms of the
second moments [see Eqs. (\ref{B19})], where the superscript $ (0)
$ corresponds to the background turbulent convection (it is a
turbulent convection without rotation, $ {\bf \Omega} = 0),$ and $
\tau (k) $ is the characteristic relaxation time of the
statistical moments. We consider the background turbulent
convection with $ \chi^{(0)}({\bf k}) = 0 .$

We assume that the characteristic times of variation of the second
moments $ f({\bf k}) ,$ $ \chi_{p}({\bf k}) , \ldots ,$ $
\Theta({\bf k}) $ are substantially larger than the correlation
time $ \tau(k) $ for all turbulence scales. This allows us to get
a stationary solution of Eqs. (\ref{EB13})-(\ref{EB18}):
\begin{eqnarray}
f({\bf k}) &=& f^{(0)}({\bf k}) - 2 \psi_{_{\Omega}}^{2}
[\mu_1({\bf k})
\nonumber\\
& & + \tau(k) g_{\perp}({\bf k}) \Phi_{R}({\bf k})] \;,
\label{B20} \\
\chi_{_{R}}({\bf k}) &=& - \psi_{_{\Lambda}} \mu_{1}({\bf k}) + k
\psi_{_{\Omega}} \psi_{_{R}} \mu_{2}({\bf k}) \;,
\label{B20B} \\
\chi_{_{I}}({\bf k}) &=& k \psi_{_{\Omega}} \mu_1({\bf k}) \;,
\label{B22} \\
\Phi_{R}({\bf k}) &=& {\Phi^{(0)}({\bf k}) \over 1 +
\psi_{_{\Omega}}^{2}} \;,
\label{B22B} \\
\Phi_{I}({\bf k}) &=& - {\psi_{_{R}} \Phi^{(0)}({\bf k}) \over (1
+ \psi_{_{\Omega}}^{2})^{2} } \;,
\label{B24} \\
F_{R}({\bf k}) &=&  k \psi_{_{\Omega}} \Phi_{I}({\bf k}) -
\psi_{_{\Lambda}} \Phi_{R}({\bf k})  \;,
\label{B24B} \\
F_{I}({\bf k}) &=& F_{I}^{(0)}({\bf k}) - k \psi_{_{\Omega}}
\Phi_{R}({\bf k}) \;,
\label{B26} \\
G({\bf k}) &=& G^{(0)}({\bf k}) + 2 (k \psi_{_{\Omega}})^{2}
\mu_1({\bf k}) \;,
\label{B26B} \\
\Theta({\bf k}) &=& \Theta^{(0)}({\bf k}) \;,
\label{B28}
\end{eqnarray}
where we changed $ \tau \psi_{_{R}} \to \psi_{_{R}} ,$ $ \quad
\tau \psi_{_{\Omega}} \to \psi_{_{\Omega}} ,$ $ \quad \tau
\psi_{_{\Lambda}} \to \psi_{_{\Lambda}} ,$
\begin{eqnarray*}
\mu_{1}({\bf k}) &=& - {1 \over 1 + 4 \psi_{_{\Omega}}^{2}}
[\varepsilon f^{(0)}({\bf k})
\nonumber\\
& & - \tau(k) g_{\perp}({\bf k}) \Phi_{R}({\bf k}) (1 - 2
\psi_{_{\Omega}}^{2})] \;,
\\
\mu_{2}({\bf k}) &=& \mu_{1}({\bf k}) - {\tau(k) g_{\perp}({\bf
k}) \Phi_{R}({\bf k}) \over 1 + \psi_{_{\Omega}}^{2}} \;,
\end{eqnarray*}
and $ f^{(0)}({\bf k}) - G^{(0)}({\bf k}) / k^{2} \equiv -
\varepsilon f^{(0)}({\bf k}) .$ Here we neglected the terms $ \sim
O[(\Lambda l_{0})^{2}] .$ We will show below that the first term
in Eq.~(\ref{B20B}), $ \chi_{_{R}}^{(1)}({\bf k}) = -
\psi_{_{\Lambda}} \mu_{1}({\bf k}) ,$ contributes to the $ \alpha
$-effect, whereas the second term in Eq.~(\ref{B20B}), $
\chi_{_{R}}^{(2)}({\bf k}) = k \psi_{_{\Omega}} \psi_{_{R}}
\mu_{2}({\bf k}) ,$ contributes to the additional effective drift
velocity. Thus, Eqs. (\ref{B20})-(\ref{B28}) describe a
modification of turbulent convection by rotation.

\subsection{The correlation tensor of velocity field}

The functions $ f({\bf k}) ,$ $ G({\bf k}) $ and $ \chi({\bf k}) $
determine the correlation tensor $ f_{ij}({\bf k}) \equiv \langle
v_i ({\bf k}) v_j(-{\bf k}) \rangle $:
\begin{eqnarray}
f_{ij}({\bf k}) &=& f_{ij}^{(a)}({\bf k}) + f_{ij}^{(b)}({\bf k})
\;,
\label{KC1} \\
f_{ij}^{(a)}({\bf k}) &=& \biggl({k_{\perp} \over k} \biggr)^{2}
\biggl\{f({\bf k}) P_{ij}(k)
\nonumber \\
& & - \biggl(f({\bf k}) - {1 \over k^{2}} G({\bf k})\biggr)
P_{ij}^{(\perp)}(k_{\perp})
\nonumber \\
& & + (i/2k^2) (k_{j} \Lambda_{i} - k_{i} \Lambda_{j}) f({\bf k})
\biggr\} \;,
\label{C1} \\
f_{ij}^{(b)}({\bf k}) &=& (1 / 2 k_{\perp}^{4}) \{ [i ({\bf k} \cdot {\bf e})
B_{ij}^{(M)} - (\Lambda / 2) B_{ij}^{(P)}
\nonumber \\
& & - i 2 k_{\perp}^{2} \varepsilon_{ijp} k_{p}] \chi_{_{R}}({\bf
k}) + [({\bf k} \cdot {\bf e}) B_{ij}^{(P)}
\nonumber \\
& & - (i \Lambda / 2) B_{ij}^{(M)} - 2 k_{\perp}^{2} A_{ij}^{(P)}]
\chi_{_{I}}({\bf k}) \} \;, \label{C2}
\end{eqnarray}
and $ P_{ij}(k) = \delta_{ij} - k_{ij} ,$ $ \quad k_{ij} = k_{i}
k_{j} / k^{2} ,$ $ \quad {\bf k} = {\bf k}_{\perp} + ({\bf k}
\cdot {\bf e}) {\bf e} ,$ $ \quad P_{ij}^{(\perp)}(k_{\perp}) =
\delta_{ij} - k^{\perp}_{ij} - e_{ij} ,$ $ \quad k^{\perp}_{ij} =
({\bf k}_{\perp})_{i} ({\bf k}_{\perp})_{j} / k_{\perp}^{2} ,$ $
\quad e_{ij} = e_{i} e_{j} ,$ and $ A_{ij}^{(P)} = ({\bf
k}_{\perp} {\bf \times} {\bf e})_{i} e_{j} + ({\bf k}_{\perp} {\bf
\times} {\bf e})_{j} e_{i} ,$ $ B_{ij}^{(P)} = ({\bf k}_{\perp}
{\bf \times} {\bf e})_{i} ({\bf k}_{\perp})_{j} + ({\bf k}_{\perp}
{\bf \times} {\bf e})_{j} ({\bf k}_{\perp})_{i} $ and $
B_{ij}^{(M)} = ({\bf k}_{\perp} {\bf \times} {\bf e})_{i} ({\bf
k}_{\perp})_{j} - ({\bf k}_{\perp} {\bf \times} {\bf e})_{j} ({\bf
k}_{\perp})_{i} .$ For the derivation of Eqs.~(\ref{C1}) and
(\ref{C2}) the velocity $ {\bf v}_{\perp} $ is written as a sum of
the vortical and the potential components, {\em i.e.,} $ {\bf
v}_{\perp} = \bec{\nabla} {\bf \times} (C {\bf e}) +
\bec{\nabla}_{\perp} \tilde \varphi ,$ where $ {\bf v} = {\bf
v}_{\perp} + v_{z} {\bf e} ,$ $\quad  w = - \Delta_{\perp} C ,$ $
\quad \Delta_{\perp} \tilde \varphi = \Lambda v_{z} / 2 -
\partial v_{z} / \partial z ,$ $ \quad \bec{\nabla} \cdot {\bf v} =
(\Lambda / 2) ({\bf v} \cdot {\bf e}) ,$ $ \quad
\bec{\nabla}_{\perp} = \bec{\nabla} - {\bf e} ({\bf e} \cdot
\bec{\nabla}) .$ We also used the identities $ ({\bf k}_{\perp}
{\bf \times} {\bf e})_{i} ({\bf k}_{\perp} {\bf \times} {\bf
e})_{j} =  k_{\perp}^{2} P_{ij}^{(\perp)}(k_{\perp}) $ and $ ({\bf
k} \cdot {\bf e}) B_{ij}^{(M)} - k_{\perp}^{2} A_{ij}^{(M)} =
k_{\perp}^{2} \varepsilon_{ijp} k_{p} $ (see, {\em e.g.,}
\cite{EKR98}). In Eq.~(\ref{C2}) we neglected the terms $ \sim
O[(\Lambda l_{0})^{2}] .$ We will use Eqs. (\ref{C1}) and
(\ref{C2}) for the calculation of the hydrodynamic helicity and
the $ \alpha $ effect.

\subsection{The hydrodynamic helicity}

Now we find the dependence of the hydrodynamic helicity $ \chi^{(v)} =
\langle {\bf u} \cdot (\bec{\nabla} {\bf \times} {\bf u}) \rangle $
on the rate of rotation and anisotropy of turbulence.
In $ {\bf k} $-space the hydrodynamic helicity is given by
\begin{eqnarray}
\chi^{(v)}({\bf k}) &\equiv& - i \, \varepsilon_{inm} k_{i}
f_{mn}({\bf k}) \exp(\Lambda z)
\nonumber \\
&=& (1 + k^{2} / k_{\perp}^{2}) \chi_{_{R}}({\bf k}) \exp(\Lambda
z) \;, \label{D2}
\end{eqnarray}
where we used Eqs. (\ref{C1}) and (\ref{C2}). The function $
\exp(\Lambda z) $ in Eq.~(\ref{D2}) implies that we used the
transformation $ {\bf u} = \exp(\Lambda z / 2) {\bf v} .$
Equation~(\ref{D2}) can be rewritten as
\begin{eqnarray}
\chi^{(v)}({\bf k}) = \exp(\Lambda z) [\chi_{1}({\bf k}) +
\chi_{2}({\bf k})] \;,
\label{D2D}
\end{eqnarray}
where
\begin{eqnarray}
\chi_{1}({\bf k}) &=& - \psi_{_{\Lambda}} \mu_{1}({\bf k}) (1 +
k^{2} / k_{\perp}^{2}) \;,
\label{D6} \\
\chi_{2}({\bf k}) &=& - 4 \tau^{2} \Omega^{2} \Lambda
(\bec{\hat \omega} \times {\bf e})_{m} \hat \omega_{n} k_{mn}
\mu_{2}({\bf k}) (1
\nonumber \\
& & + k^{2} / k_{\perp}^{2}) \;, \label{D5}
\end{eqnarray}
where $ \bec{\hat \omega} = {\bf \Omega} / \Omega $ and we used
the identity $ k \psi_{_{\Omega}} \psi_{_{R}} = - 4 \tau^{2}
\Omega^{2} \Lambda (\bec{\hat \omega} \times {\bf e})_{m} \hat
\omega_{n} k_{mn} .$ The integration in $ {\bf k} $-space in $
\chi_{1,2} = \int \chi_{1,2}({\bf k}) \,d {\bf k} $ yields
\begin{eqnarray}
\chi_{1} &=& - {1 \over 12 \delta_\ast} \biggl({l_{0}^{2} \Omega
\over L_{\rho} \tau_{_{0}}} \biggr) \sin \phi_l \, \{ (\sigma + 3)
\phi_{_{1}}\{I_{mm}^{(2)}\}
\nonumber\\
& & + (7\sigma - 9) \phi_{_{1}}\{M_{mm}^{(2)}\} - 3 (\sigma - 1)
\phi_{_{1}}\{e_{mn} M_{mn}^{(2)} \}
\nonumber\\
& & + (\lambda /2) [M_{mm}^{(1)}(2 \omega) - 2 I_{mm}^{(1)}(2
\omega)] \} \;,
\label{D8} \\
\chi_{2} &\propto& (\bec{\hat \omega} \times {\bf e})_{m} \hat
\omega_{n} [I_{mn}^{(p)}(\omega) - M_{mn}^{(p)}(\omega)] = 0 \;,
\label{D7}
\end{eqnarray}
where $ \omega = 4 \tau_{_{0}} \Omega ,$ $ \,  u_{0}^{2} = 2 g
\tau_{_{0}} \Phi^{\ast}_z \delta_\ast , $ $ \, \sin \phi_l =
\bec{\hat \omega} \cdot {\bf e} ,$ $ \, l_{0} = u_{0} \tau_{_{0}}
,$ $ \, \phi_{_{1}}\{X\} = 2 X(2\omega) - X(\omega) ,$ e.g., $ \,
\phi_{_{1}}\{I_{mn}^{(2)}\} = 2 I_{mn}^{(2)}(2\omega) -
I_{mn}^{(2)}(\omega) ,$
\begin{eqnarray}
I_{ij}^{(p)}(\omega) &=& (6 / \pi \omega^{p+1}) \int_{0}^{\omega}
y^{p} \bar I_{ij}(y^{2}) \,d y  \;,
\label{SD4}\\
M_{ij}^{(p)}(\omega) &=& (6 / \pi \omega^{p+1}) \int_{0}^{\omega}
y^{p} \bar M_{ij}(y^{2}) \,d y  \;, \label{D4}
\end{eqnarray}
$ \bar M_{ij}(y) = e_{mn} \bar I_{ijmn}(y) ,$ $ \, \bar I_{ij}(z)
$ and $ \bar I_{ijmn}(z) $ are determined by Eqs.~(\ref{AP1}) and
(\ref{AP2}) in Appendix B, and the exponent $ p = 1, 2, 3, 4 $ is
determined by $ \tau^p $ in the expressions for the the
hydrodynamic helicity, the $ \alpha $ effect and the effective
drift velocity (see below). For example, $ p = 1, 2 $ in
Eq.~(\ref{D8}). Equation (\ref{D8}) yields the angular velocity
dependence of the hydrodynamic helicity $ \chi^{(v)} $ which is
given by Eq.~(\ref{D10}).

\subsection{The electromotive force}

In order to derive equation for the $ \alpha $-tensor we introduce
the electromotive force
\begin{eqnarray}
{\cal E}_{i} = \langle {\bf u} \times {\bf b} \rangle_{i}
= \exp(\Lambda z) \varepsilon_{imn}
\int \chi^{(c)}_{mn}({\bf k}) \,d {\bf k} \;,
\label{D4D}
\end{eqnarray}
where $ \chi^{(c)}_{ij}({\bf k}) = \hat L(v_{i},h_{j}) \equiv
\langle v_{i}({\bf k}) h_{j}(-{\bf k}) \rangle $ is the
cross-helicity tensor. Using equation for $ {\bf v} = \bec{\nabla}
{\bf \times} (C {\bf e}) + \bec{\nabla}_{\perp} \tilde \varphi +
v_{z} {\bf e} $ we obtain
\begin{eqnarray}
\chi^{(c)}_{ij}({\bf k}) &=& k_{\perp}^{-2} \{ i ({\bf k} {\bf
\times} {\bf e})_{i} \xi_{j}({\bf k}) + [k_{\perp}^{2} e_{i} -
(k_{i}
\nonumber \\
& & - k_{z} e_{i})(k_{z} + i \Lambda / 2)] \chi^{(c)}_{j}({\bf k})
\} \;, \label{F3}
\end{eqnarray}
where $ \chi^{(c)}_{j}({\bf k}) = e_{i} \chi^{(c)}_{ij}({\bf k}) =
\hat L(v_{z},h_{j}) $ and $ \bec{\xi}({\bf k}) =  \hat L(w,{\bf
h}) .$ Using Eqs. (\ref{B4})-(\ref{B6}) and (\ref{F2}) we derive
equations for $ \bec{\xi}({\bf k}) ,$ $ \, \bec{\chi}^{(c)}({\bf
k}) $ and $ \bec{\zeta}({\bf k}) =  \hat L(s,{\bf h}) $:
\begin{eqnarray}
{\partial \bec{\xi} \over \partial t} &=& {\bf I}^{(w)} + (ik
\psi_{_{\Omega}} - \psi_{_{\Lambda}}) \bec{\chi}^{(c)}({\bf k}) +
\bec{\xi}_{_{N}} \;,
\label{F4} \\
{\partial \bec{\chi}^{(c)} \over \partial t} &=& {\bf I}^{(v)} +
k^{-2} (ik \psi_{_{\Omega}} + \psi_{_{\Lambda}}) \bec{\xi}({\bf
k}) + i \psi_{_{R}} \bec{\chi}^{(c)}
\nonumber\\
& & + g_{\perp}({\bf k}) \bec{\zeta} + \bec{\chi}^{(c)}_{_{N}} \;,
\label{F5}\\
{\partial \bec{\zeta} \over \partial t} &=& {\bf I}^{(s)} +
\bec{\zeta}_{_{N}} \;, \label{RF4}
\end{eqnarray}
where $ \bec{\xi}_{_{N}} ,$ $ \, \bec{\chi}^{(c)}_{_{N}} $ and $
\, \bec{\zeta}_{_{N}} $ are the third moments:
\begin{eqnarray*}
\bec{\xi}_{_{N}}({\bf k}) &=& \hat L(W_{N},{\bf h}) + \hat
L(w,{\bf H}_{N}) \;,
\\
\bec{\chi}^{(c)}_{_{N}}({\bf k}) &=& \hat L(V_{N},{\bf h}) + \hat
L(v_z,{\bf H}_{N}) \;,
\\
\bec{\zeta}_{_{N}}({\bf k}) &=& \hat L(S_{N},{\bf h}) + \hat
L(s,{\bf H}_{N}) \;,
\end{eqnarray*}
and
\begin{eqnarray}
{\bf H}_{N} &=& \sqrt{\rho_0} \, \, \bec{\nabla} {\bf \times}
({\bf u} {\bf \times} {\bf b} - \bec{\cal E} - \eta \,
\bec{\nabla} {\bf \times} {\bf b}) \;, \label{RD4}
\end{eqnarray}
$\psi_{_{\Lambda}} = {\bf \Omega} \cdot {\bf \Lambda} ,$ $ \quad
\psi_{_{\Omega}} = 2  ({\bf \Omega} \cdot {\bf k}) / k ,$ $ \quad
\psi_{_{R}} = 2 \Lambda \Omega_{x} k_{y} / k^{2} ,$ and
\begin{eqnarray}
I_j^{(v)} &=& \hat L\biggl(v_{z}, {\partial h_j \over \partial t}
\biggr) = - i \, ({\bf B} \cdot \tilde {\bf k}) e_i f_{ij}({\bf
k})
\nonumber\\
& & - \Lambda f({\bf k}) \, B_j \;,
\label{RS1}\\
{\bf I}^{(w)} &=& \hat L\biggl(w, {\partial {\bf h} \over \partial
t} \biggr) = - i \, ({\bf B} \cdot \tilde {\bf k}) \,
\bec{\chi}^{(w)}({\bf k})
\nonumber\\
& & - \Lambda \chi({\bf k}) \, {\bf B} \;,
\label{RS2}\\
{\bf I}^{(s)} &=& \hat L\biggl(s, {\partial {\bf h} \over \partial
t} \biggr) = - i \, ({\bf B} \cdot \tilde {\bf k}) \, {\bf
\Phi}({\bf k})
\nonumber\\
& & - \Lambda \Phi_z({\bf k}) \, {\bf B} \;, \label{RS3}
\end{eqnarray}
$ \tilde {\bf k} = {\bf k} + (i \Lambda / 2) {\bf e} $ and $
\bec{\chi}^{(w)}({\bf k}) \equiv \hat L(w, {\bf v}) $ is given by
\begin{eqnarray}
\bec{\chi}^{(w)}({\bf k}) &=& k_{\perp}^{-2} \{\chi({\bf k}) [{\bf
e} (k^2 - i \Lambda k_z / 2) - {\bf k} (k_z
\nonumber\\
& & - i \Lambda / 2)] - i G({\bf k}) ({\bf k} {\bf \times} {\bf
e}) \} \; . \label{RS4}
\end{eqnarray}
Note that $ \chi({\bf k}) = \bec{\chi}^{(w)}({\bf k}) \cdot {\bf
e} .$ Now we use the $ \tau $-approximation and assume that the
characteristic times of variation of the second moments $
\bec{\xi} ,$ $ \, \bec{\zeta} $ and $ \bec{\chi}^{(c)} $ are
substantially larger than the correlation time $ \tau(k) $ for all
turbulence scales. This allows us to get a stationary solution of
Eqs. (\ref{F4})-(\ref{RF4}):
\begin{eqnarray}
\bec{\xi}({\bf k}) &=& \tau {\bf I}^{(w)}({\bf k}) + (i k
\psi_{_{\Omega}} - \psi_{_{\Lambda}}) \bec{\chi}^{(c)}({\bf k})
\;,
\label{F6} \\
\bec{\chi}^{(c)} &=& {\tau (1 + \psi_{_{\Omega}}^{2} + i
\psi_{_{R}}) \over (1 + \psi_{_{\Omega}}^{2})^{2}} [{\bf I}^{(v)}
\nonumber \\
& & + k^{-2} (i k \psi_{_{\Omega}} + \psi_{_{\Lambda}}) {\bf
I}^{(w)}({\bf k})]  \;, \label{F7}
\end{eqnarray}
and $ \bec{\zeta}({\bf k}) = \tau {\bf I}^{(s)} , $ where we
changed $ \tau \psi_{_{R}} \to \psi_{_{R}} ,$ $ \quad \tau
\psi_{_{\Omega}} \to \psi_{_{\Omega}} ,$ $ \quad \tau
\psi_{_{\Lambda}} \to \psi_{_{\Lambda}} .$ Now we take into
account that a general form of the electromotive force is given by
\begin{eqnarray}
{\cal E}_{i} &=& \alpha_{ij} B_{j} +
({\bf V}^{(d)} {\bf \times} {\bf B})_{i}
- \eta_{ij} (\bec{\nabla} {\bf \times} {\bf B})_{j}
- \kappa_{ijk} (\partial \hat B)_{ij}
\nonumber \\
& & - [\bec{\delta} {\bf \times}
(\bec{\nabla} {\bf \times} {\bf B})]_{i}  \equiv a_{ij} B_{j} +
b_{ijk} B_{i,j}  \;,
\label{F7F}
\end{eqnarray}
(see, {\em e.g.,} \cite{R80}), where the tensors $ \alpha_{ij} $
and $ \eta_{ij} $ describe the $ \alpha $-effect and turbulent
magnetic diffusion, respectively, $ {\bf V}^{(d)} $ is the
effective diamagnetic (or paramagnetic) velocity, $ \kappa_{ijk} $
and $ \bec{\delta} $ describe a nontrivial behavior of the mean
magnetic field in an anisotropic turbulence, $ B_{i,j} =
\nabla_{j} B_{i} ,$ $\quad (\partial \hat B)_{ij} = (1/2) (B_{i,j}
+ B_{j,i}) .$ In this study we determine only the tensor $
\alpha_{ij} $ and the velocity $ V_{k}^{(d)} .$ The calculations
of the other coefficients defining electromotive force is a
subject of a separate paper. The tensor $ a_{ij} \equiv
a_{ij}^{(S)} + a_{ij}^{(AS)} $ follows from Eqs.~(\ref{F3}),
(\ref{F6})-(\ref{F7F}), where
\begin{eqnarray}
a_{ij}^{(S)} &=& a_{ij}^{(a)} + a_{ij}^{(b)} + a_{ij}^{(c)} \;,
\label{F50}\\
a_{ij}^{(AS)} &=& a_{ij}^{(d)} + a_{ij}^{(e)} \;,
\label{F51}
\end{eqnarray}
where $ a_{ij}^{(S)} = (1/2) (a_{ij} + a_{ji}) $ and
$ a_{ij}^{(AS)} = (1/2) (a_{ij} - a_{ji}) $ are the symmetric and
antisymmetric parts of the tensor $ a_{ij} ,$ and
\begin{eqnarray}
a_{ij}^{(a)}({\bf k}) &=& k_{ij} \psi_{_{\Lambda}} s_{1}({\bf
k},z) [2 \mu_{1}({\bf k}) + f({\bf k}) - G({\bf k}) / k^{2}
\nonumber \\
& & + 2 \tau(k) g_{\perp}({\bf k}) \Phi_R({\bf k})] \;,
\label{F9} \\
a_{ij}^{(b)}({\bf k}) &=& (e_{i} k_{j} + e_{j} k_{i}) s_{2}({\bf
k},z) [\psi_{_{\Omega}} G({\bf k}) - k \chi_{_{I}}({\bf k})
\nonumber \\
& & + k \tau(k) g_{\perp}({\bf k}) F_I({\bf k})] \; .
\label{F10} \\
a_{ij}^{(c)}({\bf k}) &=& \biggl[\biggl( {f({\bf k}) + G({\bf k})
/ k^{2} + 2 \tau(k) g_{\perp}({\bf k}) \Phi_R({\bf k}) \over 1 +
\psi_{_{\Omega}}^{2}}
\nonumber \\
&& - 2 \mu_{2}({\bf k}) \biggr)k \psi_{_{\Omega}}  -
\chi_{_{I}}({\bf k}) \biggr] k_{ij} \psi_{_{R}} s_{1}({\bf k},z)
\nonumber \\
&& + [( {\bf e} {\bf \times} {\bf k})_{i} k_{j} + ({\bf e} {\bf
\times} {\bf k})_{j} k_{i}] s_{2}({\bf k},z) [k f({\bf k})
\nonumber \\
&& + k \tau(k) g_{\perp}({\bf k}) \Phi_R({\bf k}) -
\psi_{_{\Omega}} \chi_{_{I}} ({\bf k})] \;,
\label{FF9} \\
a_{ij}^{(d)}({\bf k}) &=& [2 k^{2} \varepsilon_{ijm} e_{n}
P_{mn}(k) + ({\bf e} {\bf \times } {\bf k})_{i} k_{j}
\nonumber \\
& & - ({\bf e} {\bf \times } {\bf k})_{j} k_{i}] s_{2}({\bf k},z)
[k f({\bf k})
\nonumber \\
& & - \psi_{_{\Omega}} \chi_{_{I}} ({\bf k}) + k \tau(k)
g_{\perp}({\bf k}) \Phi_R({\bf k})] \;,
\label{FF10} \\
a_{ij}^{(e)}({\bf k}) &=& k [2 \varepsilon_{ijm} ({\bf e} {\bf
\times } {\bf k})_{m} - e_{i} k_{j}
\nonumber \\
& & + e_{j} k_{i}] s_{2}({\bf k},z) \{ k \psi_{_{\Omega}} [f({\bf
k})
\nonumber \\
& & + \tau(k) g_{\perp}({\bf k}) \Phi_R({\bf k})] +
\chi_{_{I}}({\bf k}) \} \;, \label{FF11}
\end{eqnarray}
and $ s_{1}({\bf k},z) = \exp(\Lambda z) \tau(k) (k /
k_{\perp})^{2} / (1 + \psi_{_{\Omega}}^{2}) ,$ $ \, s_{2}({\bf
k},z) = (\Lambda / 2 k^{3}) s_{1}({\bf k},z) .$  Here we used that
\begin{eqnarray}
\bec{\cal E} &=& k_{\perp}^{-2} [ i \bec{\xi} {\bf \times} ({\bf
e} {\bf \times } {\bf k}) + ({\bf e} {\bf \times}
\bec{\chi}^{(c)}) (k^2 + i \Lambda k_z / 2)
\nonumber\\
& & - ({\bf k} {\bf \times} \bec{\chi}^{(c)}) (k_z + i \Lambda /
2)] \;,
\label{MF1}\\
e_i f_{ij}({\bf k}) &=& k_{\perp}^{-2} \{[k^2 e_i P_{ij}({\bf k})
+ i (\Lambda / 2) k_i P_{ij}({\bf e})] f({\bf k})
\nonumber\\
& & + i ({\bf e} {\bf \times } {\bf k})_j [\chi_{_{R}}({\bf k}) -
i \chi_{_{I}}({\bf k})] \}  \;, \label{MF2}
\end{eqnarray}
where $ P_{ij}({\bf e}) = \delta_{ij} - e_{ij} .$ Note that $ e_i
f_{ij}({\bf k}) \not= e_i f_{ji}({\bf k}) $ because rotation
causes a nonzero helicity in the turbulent convection. Here we
also took into account that the tensor $ a_{ij} $ must be real in
$ {\bf r} $-space.

We will show that the tensors $ a_{ij}^{(a)}({\bf k}) $ and $
a_{ij}^{(b)}({\bf k}) $ contribute to the $ \alpha $-tensor, the
tensor $ a_{ij}^{(d)}({\bf k}) $ contributes to the effective
drift velocity $ {\bf V}^{(1)} ,$ the tensor $ a_{ij}^{(e)}({\bf
k}) $ contributes to the effective drift velocity $ {\bf V}^{(2)}
,$ and the tensor $ a_{ij}^{(c)}({\bf k}) $ contributes to the
effective drift velocity $ {\bf V}^{(3)} .$

\subsection{The $ \alpha $-tensor}

Now we determine the tensor $ \alpha_{ij} = a_{ij}^{(a)} +
a_{ij}^{(b)} .$ The integration in $ {\bf k} $-space yields
\begin{eqnarray}
a_{ij}^{(a)} &=& {1 \over 6 \delta_\ast} \biggl({l_{0}^{2} \Omega
\over L_{\rho}} \biggr) [\phi_{_{5}}\{I_{ij}\}
\nonumber\\
& & + 3 (\sigma - 1) \phi_{_{2}}\{M_{ij}^{(3)}\}] \sin \phi_l \;,
\label{F17} \\
a_{ij}^{(b)} &=& {1 \over 6 \delta_\ast} \biggl({l_{0}^{2} \Omega
\over L_{\rho}} \biggr) \hat P^{(b)}_{ijmn} [\phi_{_{6}}\{I_{mn}\}
\nonumber\\
& & + 3 (\sigma - 1) \phi_{_{3}}\{M_{mn}^{(3)}\}] \; .
\label{FF17}
\end{eqnarray}
where Eqs.~(\ref{F9}) and (\ref{F10}) for $ a_{ij}^{(a)}({\bf k})
$ and $ a_{ij}^{(b)}({\bf k}) ,$ respectively, $ \hat
P^{(b)}_{ijmn} = \hat \omega_{m} (e_{i} \delta_{nj} + e_{j}
\delta_{ni}) ,$ and hereafter we use the following functions:
\begin{eqnarray*}
\phi_{_{1}}\{X\} &=& 2 X(2\omega) - X(\omega) \;, \,
\\
\phi_{_{2}}\{X\} &=& 4 X(2\omega) - {3 \over \pi} \bar
X(\omega^{2}) \;,
\\
\phi_{_{3}}\{X\} &=& {3 \over \pi} \bar X(\omega^{2}) - 2
X(2\omega) - X(\omega) \;,
\\
\phi_{_{4}}\{X\} &=& 7 X^{(4)}(\omega) - 4 X^{(4)}(2\omega)
\\
& & - {3 \over \pi} \bar X(\omega^{2})  + {9 \omega^{2} \over \pi}
\biggl({\partial \bar X(a) \over \partial a}
\biggr)_{a=\omega^{2}} \;,
\\
\phi_{_{5}}\{X\} &=& (3 - \sigma) \phi_{_{2}}\{X^{(3)}\} -
(\lambda/2) [4 X^{(2)}(2 \omega)
\\
& & - X^{(2)}(\omega)] \;,
\\
\phi_{_{6}}\{X\} &=& (3 - \sigma) \phi_{_{3}}\{X^{(3)}\} +
(\lambda/2) [2 X^{(2)}(2 \omega)
\\
& & + \varepsilon^{-1} X^{(2)}(\omega)] \;,
\\
\phi_{_{7}}\{X\} &=& (3 - \sigma) \phi_{_{1}}\{X^{(2)}\} - \lambda
[X^{(1)}(2 \omega)
\\
& & - (1 + \varepsilon^{-1}) X^{(1)}(\omega)] \;,
\\
\phi_{_{8}}\{X\} &=& (3 - \sigma) \phi_{_{1}}\{X^{(3)}\} -
(\lambda/2) [\phi_{_{1}}\{X^{(2)}\}
\\
& & - \varepsilon^{-1} X^{(2)}(\omega)] \;,
\\
\phi_{_{9}}\{X\} &=& 4 \lambda X^{(3)}(2 \omega) - (\lambda + 2
\delta_\ast) X^{(3)}(\omega)
\\
& & + {6 \delta_\ast \over \pi} \bar X(\omega^{2})  \; .
\end{eqnarray*}
For example,
\begin{eqnarray*}
\phi_{_{3}}\{M_{mn}^{(3)}\} &=& {3 \over \pi} \bar
M_{mn}(\omega^{2}) - 2 M_{mn}^{(3)}(2\omega) -
M_{mn}^{(3)}(\omega)  \;,
\\
\phi_{_{6}}\{I_{mn}\} &=& (3 - \sigma) \phi_{_{3}}\{I_{mn}^{(3)}\}
+ (\lambda/2) [2 I_{mn}^{(2)}(2 \omega)
\\
& & + \varepsilon^{-1} I_{mn}^{(2)}(\omega)] \;,
\end{eqnarray*}
the functions $ I_{mn}^{(2)}(\omega) $ and $ M_{mn}^{(3)}(\omega)
$ are determined by Eqs.~(\ref{SD4}) and (\ref{D4}), $ \bar
M_{ij}(y) = e_{mn} \bar I_{ijmn}(y) ,$ and the functions $ \bar
I_{ij}(z) $ and $ \bar I_{ijmn}(z) $ are determined by
Eqs.~(\ref{AP1}) and (\ref{AP2}) in Appendix B. Now we use the
following identities
\begin{eqnarray*}
\hat P^{(b)}_{ijmn} \bar I_{mn} &=& (e_{i} \hat \omega_{j} + e_{j}
\hat \omega_{i}) \bar L_1 \;,
\\
\hat P^{(b)}_{ijmn} \bar M_{mn} &=& (e_{i} \hat \omega_{j} + e_{j}
\hat \omega_{i}) (\bar L_3 + \bar L_2 \sin^2 \phi_l)
\\
& & + 4 \, e_{ij} \, \bar L_3 \, \sin \phi_l  \,,
\end{eqnarray*}
where $ \bar L_k $ are determined by Eqs.~(\ref{M25}) in Appendix
C. Thus, the $ \alpha $-tensor, $ \alpha_{ij} \equiv a_{ij}^{(a)}
+ a_{ij}^{(b)} ,$ is given by Eq.~(\ref{F12}). For a slow rotation
$ (\omega \ll 1) $ the tensor $ \alpha_{ij} $ is given by
\begin{eqnarray}
\alpha_{ij} &\approx& \alpha \delta_{ij} - {4 \over 5}
\biggl({l_{0}^{2} \Omega \over L_{\rho}} \biggr) \biggl(1 -
{\sigma \over 6}
\nonumber \\
& & - {5 \lambda \over 9} [1 + (2 \varepsilon)^{-1}] \biggr)
(e_{i} \hat \omega_{j} + e_{j} \hat \omega_{i}) \;, \label{FF14}
\end{eqnarray}
and for $ \omega \gg 1 $ it is given by
\begin{eqnarray}
\alpha_{ij}  & \approx & \alpha \delta_{ij} + {\pi \over 16}
\biggl({l_{0} u_{0} \over L_{\rho}} \biggr) \biggl[\omega_{ij}
\sin \phi_l \biggl(\lambda + {\sigma \over 6}
\nonumber \\
& & + {3 \over 2} (\sigma - 1) \sin^2 \phi_l - {3 \over 2} \biggr)
- (e_{i} \hat \omega_{j} + e_{j} \hat \omega_{i}) (1
\nonumber \\
& & + 2 \cos^2 \phi_l) \biggr] \; .
\label{FF15}
\end{eqnarray}

\subsection{The effective drift velocity}

Now we determine the effective drift velocity $ V_{m}^{(d)} \equiv
V_{m}^{(1)} + V_{m}^{(2)} ,$ where
\begin{eqnarray}
V_{m}^{(1)} &=& - (1/2) \varepsilon_{mij} a_{ij}^{(d)} = \hat
P^{(d)}_{mij} \tilde a_{ij}^{(d)} \;,
\label{EE1}\\
V_{m}^{(2)} &=& - (1/2) \varepsilon_{mij} a_{ij}^{(e)} \;,
\label{E1}
\end{eqnarray}
and
\begin{eqnarray}
a_{ij}^{(d)} &=& e_p (2 \varepsilon_{ijp}  \delta_{mn} +
\varepsilon_{ipn} \delta_{jm} - \varepsilon_{jpn} \delta_{im}
\nonumber\\
& & - 2 \varepsilon_{ijm} \delta_{np}) \tilde a_{mn}^{(d)} \;,
\label{RE3}\\
\tilde a_{ij}^{(d)} &=& {1 \over 48 \delta_\ast} \biggl({l_{0}
u_{0} \over L_{\rho}} \biggr) [\phi_{_{7}}\{I_{ij}\}
\nonumber\\
& & + 3 (\sigma - 1) \phi_{_{1}}\{M_{ij}^{(2)}\}]  \;,
\label{RE1}\\
a_{ij}^{(e)} &=& {1 \over 6 \delta_\ast} \biggl({l_{0}^{2} \Omega
\over L_{\rho}} \biggr) \hat P^{(e)}_{ijmn} [\phi_{_{8}}\{I_{mn}\}
\nonumber\\
& & + 3 (\sigma - 1) \phi_{_{1}}\{M_{mn}^{(3)}\}]  \;,
\label{RE4}
\end{eqnarray}
$ \hat P^{(d)}_{imn} = 2 e_n \delta_{mi} - e_m \delta_{ni} - e_i
\delta_{mn} ,$ $ \, \hat P^{(e)}_{ijmn} = (e_i \delta_{jn} - e_j
\delta_{in}) \hat \omega_{m} .$ For the integration in $ {\bf k}
$-space we used Eqs.~(\ref{FF10}) and (\ref{FF11}) for $
a_{ij}^{(d)}({\bf k}) $ and $ a_{ij}^{(e)}({\bf k}) ,$
respectively. Using the following identities:
\begin{eqnarray*}
\hat P^{(d)}_{imn} \bar I_{mn} &=& - e_i \bar L_{4} + \hat
\omega_i \sin \phi_l \, \bar A_{2} \;,
\\
\hat P^{(d)}_{imn} \bar M_{mn} &=& e_i [3 \bar C_{1} - \bar A_{1}
+ (3 \bar C_{3} - \bar A_{2}) \sin^2 \phi_l]
\\
& & + \hat \omega_i \sin \phi_l [3 \bar C_{3} + \bar C_{2} \sin^2
\phi_l] \;,
\\
\varepsilon_{kij} \hat P^{(e)}_{ijmn} \bar I_{mn} &=& - 2 \bar
L_{1} (\bec{\hat \omega} {\bf \times} {\bf e})_{k} \;,
\\
\varepsilon_{kij} \hat P^{(e)}_{ijmn} \bar M_{mn} &=& - 2 (\bar
L_{3} + \bar L_{2} \sin^2 \phi_l) (\bec{\hat \omega} {\bf \times}
{\bf e})_{k} \;
\end{eqnarray*}
in Eqs. (\ref{EE1})-(\ref{RE4}), we obtain the effective drift
velocities $ V_{i}^{(1)} $ and $ V_{i}^{(2)} $ which are given by
Eqs.~(\ref{E4}) and (\ref{E15}). Here $ \bar M_{ij}(y) = e_{mn}
\bar I_{ijmn}(y) ,$ $ \, \bar I_{ij}(z) $ and $ \bar I_{ijmn}(z) $
are determined by Eqs.~(\ref{AP1}) and (\ref{AP2}) in Appendix B,
$ \bar L_k $ are determined by Eqs.~(\ref{M25}) in Appendix C.

The electromotive force has a term $ a_{ij}^{(c)} B_{j} $ which
for an axisymmetric case contributes only to an additional
effective drift velocity, $ {\bf V}^{(3)} ,$ of the mean magnetic
field, {\em i.e.,} $ a_{ij}^{(c)} B_{j} = [{\bf V}^{(3)} {\bf
\times} ({\bf B}_{p} - {\bf B}_{T})]_{i} ,$ where $ {\bf B} = {\bf
B}_{T} + {\bf B}_{p} $ is the mean magnetic field with the
toroidal $ ({\bf B}_{T}) $ and poloidal $ ({\bf B}_{p}) $
components, the tensor $ a_{ij}^{(c)}({\bf k}) $ is determined by
Eq.~(\ref{FF9}). Integration in $ {\bf k} $-space, $ a_{ij}^{(c)}
= \int a_{ij}^{(c)}({\bf k}) \,d {\bf k} ,$ yields
\begin{eqnarray*}
a_{ij}^{(c)} &=& - {1 \over 48 \delta_\ast} \biggl({l_{0} u_{0}
\over L_{\rho}} \biggr) \{2 \omega^2 (\bec{\hat \omega} {\bf
\times} {\bf e})_{m} \hat \omega_n [\phi_{_{9}}\{I_{ijmn}\}
\\
&& + 2 (3 - \sigma) \phi_{_{4}}\{I_{ijmn}\} + 6 (\sigma - 1)
\phi_{_{4}}\{J_{ijmn}\}]
\\
&& -  \hat P^{(c)}_{ijmn} [3 (\sigma - 1)
\phi_{_{1}}\{M_{mn}^{(2)}\} + \phi_{_{7}}\{I_{mn}\}]\}  \;,
\end{eqnarray*}
where $ \hat P^{(c)}_{ijmn} = (\varepsilon_{ipn} \delta_{jm} +
\varepsilon_{jpn} \delta_{im}) e_p .$ In order to determine the
effective drift velocity $ V_{k}^{(3)} $ we use the following
identities:
\begin{eqnarray*}
(c_{i} q_{j} + c_{j} q_{i}) B_{j} &=& [({\bf q} {\bf \times} {\bf c})
{\bf \times} ({\bf B}_{p} - {\bf B}_{T})]_{i} \;,
\\
\hat P^{(c)}_{ijmn} \bar I_{mn} &=& - \bar A_{2} (c_{i} \hat
\omega_{j} + c_{j} \hat \omega_{i}) \;,
\\
\hat P^{(c)}_{ijmn} \bar M_{mn} &=& - (\bar C_{3} + \bar C_{2}
\sin^2 \phi_l) (c_{i} \hat \omega_{j} + c_{j} \hat \omega_{i})
\\
& & - 2 \bar C_{3} \sin \phi_l \, (c_{i} e_{j} + c_{j} e_{i}) \;,
\\
c_{m} \hat \omega_{n} \bar I_{ijmn} &=& \bar L_{3}
(c_{i} \hat \omega_{j} + c_{j} \hat \omega_{i}) \;,
\\
c_{m} \hat \omega_{n} \bar J_{ijmn} &=& (\bar L_{5} + \bar L_{6}
\sin^2 \phi_l) (c_{i} \hat \omega_{j} + c_{j} \hat \omega_{i})
\\
&& + \bar D_{4} \sin \phi_l \, (c_{i} e_{j} + c_{j} e_{i}) \;,
\end{eqnarray*}
where $ c_{i} = (\bec{\hat \omega} {\bf \times} {\bf e})_{i} ,$ $
\, {\bf q} = \bec{\hat \omega} $ or $ {\bf q} = {\bf e} ,$ $ \,
\bar M_{ij} = \bar J_{ijmm} ,$
\begin{eqnarray}
I_{ijmn}^{(p)}(\omega) &=& (6 / \pi \omega^{p+1})
\int_{0}^{\omega} y^{p} \bar I_{ijmn}(y^{2}) \,d y  \;,
\label{MD3} \\
J_{ijmn}^{(p)}(\omega) &=& (6 / \pi \omega^{p+1})
\int_{0}^{\omega} y^{p} \bar J_{ijmn}(y^{2}) \,d y \;,
\label{D3}
\end{eqnarray}
and we used Eqs. (\ref{AP1})-(\ref{M10}) in Appendix B and Eqs.
(\ref{MA26})-(\ref{M25}) in Appendix C. Thus, the effective drift
velocity $ {\bf V}^{(3)} $ is given by Eq.~(\ref{E8}).

\section{The identities used for the integration in $ {\bf k} $--space}

To integrate over the angles in $ {\bf k} $--space we used the
following identities:
\begin{eqnarray}
\bar I_{ij}(a) = \int {k_{ij} \sin \theta \over 1 + a \cos^{2}
\theta} \,d\theta \,d\varphi =  \bar A_{1} \delta_{ij} + \bar
A_{2} \, \omega_{ij} \;, \label{AP1}
\end{eqnarray}
\begin{widetext}
\begin{eqnarray}
\bar I_{ijmn}(a) &=& \int {k_{ijmn} \sin \theta \over 1 + a
\cos^{2} \theta} \,d\theta \,d\varphi = \bar C_{1} (\delta_{ij}
\delta_{mn} + \delta_{im} \delta_{jn} + \delta_{in} \delta_{jm}) +
\bar C_{2} \, \omega_{ijmn}
\nonumber\\
& &+ \bar C_{3} (\delta_{ij} \omega_{mn}
+ \delta_{im} \omega_{jn} + \delta_{in} \omega_{jm} +
\delta_{jm} \omega_{in} + \delta_{jn} \omega_{im} + \delta_{mn} \omega_{ij})
\;,
\label{AP2} \\
\bar J_{ijmn}(a) &=& e_{pq} \int {k_{ijmnpq} \sin \theta \over 1 +
a \cos^{2} \theta} \,d\theta \,d\varphi = [\bar D_{1} + {1\over 3}
\bar D_{3} (\bec{\hat \omega} \cdot {\bf e})^{2}] (\delta_{ij}
\delta_{mn} + \delta_{im} \delta_{jn} + \delta_{in} \delta_{jm})
\nonumber\\
& & + [\bar D_{2} + \bar D_{7} (\bec{\hat \omega} \cdot {\bf e})^{2}]
\omega_{ijmn} + \bar D_{5} (\bec{\hat \omega} \cdot {\bf e})
(\omega_{ijm} e_{n} + \omega_{ijn} e_{m} + \omega_{jmn} e_{i} +
\omega_{imn} e_{j})
\nonumber\\
& & + [\bar D_{3} + \bar D_{6} (\bec{\hat \omega} \cdot {\bf e})^{2}]
(\delta_{ij} \omega_{mn} + \delta_{im} \omega_{jn} + \delta_{in} \omega_{jm} +
\delta_{jm} \omega_{in} + \delta_{jn} \omega_{im} + \delta_{mn} \omega_{ij})
\nonumber\\
& & + \bar D_{4} (\delta_{ij} e_{mn}
+ \delta_{im} e_{jn} + \delta_{in} e_{jm}
+ \delta_{jm} e_{in} + \delta_{jn} e_{im} + \delta_{mn} e_{ij})
\nonumber\\
& & - {3\over 4} \bar D_{3} (e_{ij} \omega_{mn} + e_{im}
\omega_{jn} + e_{in} \omega_{jm} + e_{jm} \omega_{in} + e_{jn}
\omega_{im} + e_{mn} \omega_{ij}) \;,
\label{AP3} \\
\bar H_{ijmn}(a) &=& \int {k_{ijmn} \sin \theta \over (1 + a
\cos^{2} \theta)^{2} } \,d\theta \,d\varphi = - \biggl( {\partial
\over \partial b } \int {k_{ijmn} \sin \theta \over b + a \cos^{2}
\theta} \,d\theta \,d\varphi \biggr)_{b=1}
\nonumber\\
&=& \bar I_{ijmn}(a) + a {\partial \over \partial a} \bar
I_{ijmn}(a) \;,
\label{AP4}\\
\bar G_{ijmn}(a) &=& \int {k_{ijmn} \sin \theta \over (1 + a
\cos^{2} \theta)^{3} } \,d\theta \,d\varphi = \bar H_{ijmn}(a) +
{a \over 2} {\partial \over \partial a} \bar H_{ijmn}(a) \;,
\label{AAP4}\\
\bar M_{ij}(a) &=& (\bar C_{1} + \bar C_{3} \sin^2 \phi_l) \,
\delta_{ij} + (\bar C_{3} + \bar C_{2} \sin^2 \phi_l) \,
\omega_{ij} + 2 \bar C_{1} \, e_{ij}
\nonumber\\
& & + 2 \bar C_{3} \sin \phi_l \, (e_{i} \omega_{j} + e_{j}
\omega_{i}) \;, \label{MAP4}
\end{eqnarray}
\end{widetext}
\noindent
where $ \bar M_{ij}(a) \equiv e_{mn} \bar I_{ijmn}(a),$
\begin{eqnarray}
e_{ij} \bar M_{ij}(a) &=& 3 \bar C_{1} + 6 \bar C_{3} \sin^2
\phi_l + \bar C_{2} \sin^4 \phi_l \;,
\label{MBP4}\\
\bar M_{pp}(a) &\equiv& e_{ij} \bar I_{ij}(a) = \bar A_{1} + \bar
A_{2} \sin^2 \phi_l \;,
\label{MCP4}
\end{eqnarray}
and $ \omega_{ij} = \hat \omega_{i} \hat \omega_{j} ,$ $ \quad
\omega_{ijm} = \hat \omega_{i} \hat \omega_{j} \hat \omega_{j} ,$
$ \, \bar A_{1} = 5 \bar C_{1} + \bar C_{3} ,$ $ \, \bar A_{2} =
\bar C_{2} + 7 \bar C_{3} ,$ and
\begin{eqnarray}
\bar A_{1}(a) &=& \bar F(1; -1; 0; 0) \;, \,
\bar A_{2}(a)  =  \bar F(-1; 3; 0; 0) \;, \,
\nonumber\\
\bar C_{1}(a)  &=& (1/4) \bar F(1; -2; 1; 0) \;, \,
\nonumber\\
\bar C_{2}(a) &=& (1/4) \bar F(3; -30; 35; 0) \;, \,
\nonumber\\
\bar C_{3}(a)  &=& (1/4) \bar F(-1; 6; -5; 0) \;, \,
\nonumber\\
\bar C_{4}(a) &=& \bar F(0; 0; 1; -1) \;, \, \bar C_{5}(a)  = \bar
F(0; 0; -1; 3) \;, \,
\nonumber\\
\bar D_{1}(a)  &\equiv&- (1/8) (\bar C_{1} + 5 \bar C_{3} - 5 \bar C_{4})
\nonumber\\
&=& (1/8) \bar F(1; -7; 11; -5) \;,
\nonumber\\
\bar D_{2}(a) &\equiv& - (1/8) (51 \bar C_{1} + 111 \bar C_{3}
- 119 \bar C_{4})
\nonumber\\
&=& (1/8) \bar F(15; -141; 245; -119) \;,
\nonumber\\
\bar D_{3}(a) &\equiv& (3/8) (3 \bar C_{1} + 7 \bar C_{3} - 7 \bar C_{4})
\nonumber\\
&=& (3/8) \bar F(-1; 9; -15; 7) \;,
\nonumber\\
\bar D_{4}(a) &\equiv& (1/2) (\bar C_{1} + \bar C_{3} - \bar C_{4})
\nonumber\\
&=& (1/2) \bar F(0; 1; -2; 1) \;,
\nonumber\\
\bar D_{5}(a)  &\equiv& 3 \bar C_{1} + 9 \bar C_{3} - 7 \bar C_{4}
\nonumber\\
&=&  (1/2) \bar F(-3; 24; -35; 14) \;,
\nonumber\\
\bar D_{6}(a) &\equiv& (1/8) (5 \bar C_{2} + 3 \bar C_{3} + 20 \bar C_{4} -
5 \bar C_{5})
\nonumber\\
&=& (1/8) \bar F(3; -33; 65; -35) \;,
\nonumber\\
\bar D_{7}(a) &\equiv& - (1/8) (48 \bar C_{1} + 27 \bar C_{2} +
165 \bar C_{3} + 28 \bar C_{4}
\nonumber\\
& & - 35 \bar C_{5}) = (1/8) \bar F(9; -21; -105; 133) \; .
\label{M10}
\end{eqnarray}
Here
\begin{eqnarray*}
\bar F(\tilde\alpha; \tilde\beta; \tilde\gamma; \tilde\mu) = \pi
[\tilde\alpha \bar J_{0}(a) + \tilde\beta \bar J_{2}(a) +
\tilde\gamma \bar J_{4}(a) + \tilde\mu \bar J_{6}(a)] \;,
\end{eqnarray*}
\begin{eqnarray}
\bar J_{2k}(a) &\equiv& 2 \int_{0}^{1} x^{2k} / (1 + a x^{2}) \,
dx = a^{-1}[2/(2k - 1)
\nonumber\\
& & - \bar J_{2(k-1)}(a)] \;, \label{KM11}
\end{eqnarray}
and $ \bar J_{0}(a) = 2 \arctan (\sqrt{a}) \, / \, \sqrt{a} .$ In
the case of $ a \ll 1 $ these functions are given by
\begin{eqnarray*}
\bar J_{2k}(a) &\sim& {2 \over 2 k + 1} \biggl[1 - a {2 k + 1 \over 2 k + 3}
+ a^{2} {2 k + 1 \over 2 k + 5}  \biggr] \;,
\end{eqnarray*}
and for $ a \gg 1 $ they are given by $ \bar J_{2k}(a) \sim 2 /
a \, (2k - 1) $ for all integer $ k $ except for $ k = 0 $ and
$ \bar J_{0}(a) \sim \pi / \sqrt{a} - 2 / a .$
Now we introduce the following functions:
\begin{eqnarray*}
F^{(p)}(\tilde\alpha; \tilde\beta; \tilde\gamma; \tilde\mu) = (6 /
\pi \omega^{p+1}) \int_{0}^{\omega} y^{p} \bar F(\tilde\alpha;
\tilde\beta; \tilde\gamma; \tilde\mu) |_{a=y^{2}} \,d y
\\
\equiv \tilde\alpha J_{0}^{(p)}(\omega) + \tilde\beta
J_{2}^{(p)}(\omega) + \tilde\gamma J_{4}^{(p)}(\omega) + \tilde\mu
J_{6}^{(p)}(\omega) \;, \label{M12}
\end{eqnarray*}
where
\begin{eqnarray}
J_{2k}^{(p)}(\omega) = (6 / \omega^{p+1}) \int_{0}^{\omega}
y^{p} \bar J_{2k}(y^{2}) \,d y \; .
\label{M14}
\end{eqnarray}
The integration in Eq.~(\ref{M14}) yields:
\begin{eqnarray}
J_{2k}^{(p)}(\omega) = \omega^{-2} \biggl({12 \over (2 k - 1) (p -
1)} - J_{2(k-1)}^{(p-2)}(\omega) \biggr) \label{AA10}
\end{eqnarray}
for $ p \not= 1 $ and all integer $ k $ except for $ k = 0 .$
When $ p = 1 $ and $ k \not= 0 $ we get:
\begin{eqnarray}
J_{2k}^{(1)}(\omega) &=& {6 \over 2 k - 1} \biggl[ {\ln(1 + \omega^{2})
\over \omega^{2}} + (-1)^{k} {2 \over \omega^{2k}}
\biggl({\arctan (\omega) \over \omega}
\nonumber\\
& & - \sum_{m=0}^{k-1} {(-1)^{m} \omega^{2m} \over (2m+1)} \biggr) \biggr] \; .
\label{M15}
\end{eqnarray}
When $ k = 0 $ we obtain
\begin{eqnarray}
J_{0}^{(2n)}(\omega) &=& {6 \over n} \biggl[{\arctan (\omega) \over \omega}
\biggl(1 + {(-1)^{n+1} \over \omega^{2n}} \biggr)
\nonumber\\
& & + \sum_{m=1}^{n-1} {(-1)^{n+m-1} \over (2m-1) \,
\omega^{n-m+1}} \biggr] \;,
\label{M16} \\
J_{0}^{(2n+1)}(\omega) &=& {6 \over 2n+1} \biggl[2 {\arctan (\omega)
\over \omega} + {(-1)^{n+1} \over \omega^{2(n+1)}} \ln(1 + \omega^{2})
\nonumber\\
& & - n! \sum_{m=1}^{n-1} (-1)^{m} {(1 + \omega^{2})^{n-m} - 1
\over (n-m) \, m! \, (n-m)!} \biggr] \; .
\nonumber\\
\label{M17}
\end{eqnarray}
Equation~(\ref{M16}) is for all integer $ n > 1 ,$ and $
J_{0}^{(0)}(\omega) = (12 / \omega) \int_{0}^{\omega} [\arctan (y)
/ y] \,d y .$ For $ n = 0 $ and $ n = 1 $ the third term with the
sum in Eq.~(\ref{M17}) should be dropped. In order to use
Eq.~(\ref{AA10}) for $ p = 2 $ we need to know the function $
J_{2k}^{(0)}(\omega) $ which is given by
\begin{eqnarray}
J_{2k}^{(0)}(\omega) &=& {3 \over 2^{k-2}} \biggl[{\arctan (\omega)
\over \omega} \biggl(1 + {(-1)^{k+1} \over \omega^{2k}} \biggr)
\nonumber\\
& & + \sum_{m=1}^{k-1} {(-1)^{k+m} \over (2m+1) \, \omega^{2(k-m)}} \biggr] \; .
\label{M18}
\end{eqnarray}
In the case of $ \omega \ll 1 $ these functions are given by
\begin{eqnarray*}
J_{2k}^{(p)}(\omega) &\sim& {12 \over (2 k + 1) (p + 1)} \biggl[1
- \omega^{2} \, \biggl({2 k + 1 \over 2 k + 3}\biggr)
\biggl({p + 1 \over p + 3} \biggr)
\\
& & + \omega^{4} \, \biggl({2 k + 1 \over 2 k + 5}\biggr)
\biggl({p + 1 \over p + 5} \biggr) \biggr] \; .
\end{eqnarray*}
In the case of $ \omega \gg 1 $ these functions are given by
\begin{eqnarray*}
J_{2k}^{(p)}(\omega) \sim {12 \over (2 k - 1) (p - 1) \,
\omega^{2}} \;
\end{eqnarray*}
for $ p \not = 1 $ and $ k \not= 0 ;$
\begin{eqnarray*}
J_{0}^{(p)}(\omega) \sim {6 \pi \over p \, \omega} -
{12 \over (p - 1) \, \omega^{2}} \;
\end{eqnarray*}
for $ p \not = 0 $ and $ p \not= 1 ;$
\begin{eqnarray*}
J_{2k}^{(1)}(\omega) &\sim& {12 \ln \, \omega \over (2 k - 1)
\, \omega^{2}} \;
\end{eqnarray*}
for $ k \not= 0 ;$ and
\begin{eqnarray*}
J_{2k}^{(0)}(\omega) &\sim& {3 \pi \over 2^{k-2} \, \omega} \biggl[1
- {1 \over \pi \, \omega} \biggl({4k - 1 \over 2k - 1} \biggr)
\biggr] \;,
\\
J_{0}^{(1)}(\omega) &\sim& {6 \pi \over \omega} \biggl(1 - {2 \ln
\, \omega \over \pi \omega} \biggr) \;, \, J_{0}^{(0)}(\omega)
\sim {6 \pi \, \ln \, \omega \over \omega} \; .
\end{eqnarray*}
Now we introduce the following functions
\begin{eqnarray}
H_{ijmn}^{(p)}(\omega) &=& (6 / \pi \omega^{p+1}) \int_{0}^{\omega}
y^{p} \bar H_{ijmn}(y^{2}) \,d y
\nonumber\\
&=& (3 / \pi) \bar I_{ijmn}(\omega^{2}) - (p-1)
I_{ijmn}^{(p)}(\omega) / 2 \;,
\label{RP1}\\
G_{ijmn}^{(p)}(\omega) &=& (6 / \pi \omega^{p+1})
\int_{0}^{\omega} y^{p} \bar G_{ijmn}(y^{2}) \,d y
\nonumber\\
&=& \biggl({p-1 \over 2} \biggr)^{2} I_{ijmn}^{(p)}(\omega) +
\biggl({3(3-p) \over 2 \pi} \biggr) \bar I_{ijmn}(\omega^{2})
\nonumber\\
& & + {3 \omega^{2} \over \pi} \biggl({\partial \bar I_{ijmn}(a)
\over \partial a} \biggr)_{a=\omega^{2}} \;, \label{RP2}
\end{eqnarray}
which will be used for the calculation of the effective drift
velocity of the mean magnetic field. The functions $
A_{k}^{(p)}(\omega) $ can be obtained from Eqs.~(\ref{M10}) after
the change of LHS of Eqs.~(\ref{M10}) $ \bar A_{k}(a) \to
A_{k}^{(p)}(\omega) $ and of RHS of Eqs.~(\ref{M10}) $ \bar
F(\tilde\alpha; \tilde\beta; \tilde\gamma; \tilde\mu) \to
F^{(p)}(\tilde\alpha; \tilde\beta; \tilde\gamma; \tilde\mu) $ and
similarly for the functions $ C_{k}^{(p)}(\omega) $ and $
D_{k}^{(p)}(\omega) ,$ {\em e.g.,}
\begin{eqnarray}
A_{1}^{(p)}(\omega) &=& F^{(p)}(1; -1; 0; 0) \;, \,
\nonumber\\
C_{1}^{(p)}(\omega)  &=& (1/4) F^{(p)}(1; -2; 1; 0) \;, \, \ldots
\label{M26M}
\end{eqnarray}
and similarly for the other functions $ C_{k}^{(p)}(\omega) $ and
$ D_{k}^{(p)}(\omega) .$ For the calculation of the functions $
\phi_{_{4}}\{X\} $ we need to use the following identities:
\begin{eqnarray*}
\omega^{2} \biggl({\partial \bar J_{2k}(a) \over
\partial a} \biggr)_{a=\omega^{2}}
= - \biggl(\bar J_{2k}(a) + {\partial \bar J_{2(k-1)}(a) \over
\partial a} \biggr)_{a=\omega^{2}} \;,
\end{eqnarray*}
where
\begin{eqnarray*}
{\partial \bar J_{0}(a) \over \partial a} = {1 \over 2 a} \biggl(
{4 \sqrt{a} \over a + 1} - \bar J_{0}(a) \biggr) \; .
\end{eqnarray*}

\section{The functions $\Psi_{\beta}(\omega) $
and $E_{\beta}(\omega)$}

The functions $ \Psi_{k}(\omega) $ are given by
\begin{eqnarray}
\Psi_{1}(\omega) &=& 10 \sigma \phi_{_{1}}\{A_{1}^{(2)}\} +
(\sigma + 3) \phi_{_{1}}\{A_{2}^{(2)}\}
\nonumber \\
& & - (\lambda/2)  [5 A_{1}^{(1)}(2 \omega) + A_{2}^{(1)}(2
\omega)]
\nonumber \\
& & - 9 (\sigma - 1) \phi_{_{1}}\{C_{1}^{(2)}\} \;,
\nonumber \\
\Psi_{2}(\omega) &=& (7 \sigma - 9) \phi_{_{1}}\{A_{2}^{(2)}\} -
18 (\sigma - 1) \phi_{_{1}}\{C_{3}^{(2)}\}
\nonumber \\
& & + (\lambda/2) A_{2}^{(1)}(2 \omega) \;,
\nonumber \\
\Psi_{3}(\omega) &=& - 3 (\sigma - 1) \phi_{_{1}}\{C_{2}^{(2)}\}
\;,
\nonumber \\
\Psi_{4}(\omega) &=& \phi_{_{5}}\{A_{1}\} + 3 (\sigma - 1)
\phi_{_{2}}\{C_{1}^{(3)}\} \;,
\nonumber \\
\Psi_{5}(\omega) &=& 3 (\sigma - 1) \phi_{_{2}}\{C_{3}^{(3)}\}
\;,
\nonumber \\
\Psi_{6}(\omega) &=& \phi_{_{5}}\{A_{2}\} + 3 (\sigma - 1)
\phi_{_{2}}\{C_{3}^{(3)}\} \;,
\nonumber \\
\Psi_{7}(\omega) &=& 3 (\sigma - 1) \phi_{_{2}}\{C_{2}^{(3)}\} \;,
\nonumber \\
\Psi_{8}(\omega) &=& 6 (\sigma - 1) [\phi_{_{2}}\{C_{1}^{(3)}\} +
2 \phi_{_{3}}\{L_{3}^{(3)}\}]  \;,
\nonumber \\
\Psi_{9}(\omega) &=& \phi_{_{6}}\{L_{1}\} + 3 (\sigma - 1)
\phi_{_{3}}\{L_{3}^{(3)}\} \;,
\nonumber \\
\Psi_{10}(\omega) &=& 2 \Psi_{5}(\omega) + 3 (\sigma - 1)
\phi_{_{3}}\{L_{2}^{(3)}\} \; . \label{M50}
\end{eqnarray}

The functions $ E_{k}(\omega) $ are given by
\begin{eqnarray}
E_{1}(\omega) &=& 3 (\sigma - 1) \phi_{_{1}}\{L_{7}^{(2)}\} -
\phi_{_{7}}\{L_{4}\} \;,
\nonumber\\
E_{2}(\omega) &=& \phi_{_{7}}\{A_{2}\} + (3 \sigma - 1)
\phi_{_{1}}\{L_{8}^{(2)}\} \;,
\nonumber\\
E_{3}(\omega) &=& 3 (\sigma - 1) \phi_{_{1}}\{C_{2}^{(2)}\} \;,
\nonumber\\
E_{4}(\omega) &=& \phi_{_{7}}\{A_{2}\} + 9 (\sigma - 1)
\phi_{_{1}}\{C_{3}^{(2)}\} \;,
\nonumber\\
E_{5}(\omega) &=& \phi_{_{8}}\{L_{1}\} + 3 (\sigma - 1)
\phi_{_{1}}\{L_{3}^{(3)}\} \;,
\nonumber\\
E_{6}(\omega) &=& 3 (\sigma - 1) \phi_{_{1}}\{L_{2}^{(3)}\} \;,
\nonumber\\
E_{7}(\omega) &=& (1/2) [\phi_{_{7}}\{A_{2}\} + 3 (\sigma - 1)
\phi_{_{1}}\{C_{3}^{(2)}\}]
\nonumber\\
& & + \omega^{2} [\phi_{_{9}}\{L_{3}\} + 2 (3 - \sigma)
\phi_{_{4}}\{L_{3}\}
\nonumber\\
& & + 6 (\sigma - 1) \phi_{_{4}}\{L_{5}\}] \;,
\nonumber\\
E_{8}(\omega) &=& - E_{7}(\omega) - E_{9}(\omega) \;,
\nonumber\\
E_{9}(\omega) &=& - (3/2) (\sigma - 1) \phi_{_{1}}\{C_{2}^{(2)}\}
- 6 \omega^{2} (\sigma - 1) \phi_{_{4}}\{L_{6}\} \;,
\nonumber\\
E_{10}(\omega) &=& E_{1}(\omega) + 3 (\sigma - 1)
[\phi_{_{1}}\{C_{3}^{(2)}\}
\nonumber\\
& & + 2 \omega^{2} \phi_{_{4}}\{D_{4}\}] \;, \label{MA26}
\end{eqnarray}
where
\begin{eqnarray}
\bar L_{1}(a) &\equiv& \bar A_{1} + \bar A_{2} = 2 \bar F(0; 1; 0;
0) \;, \,
\nonumber\\
\bar L_{2}(a) &\equiv& \bar C_{2} + 3 \bar C_{3} = \bar F(0; -3;
5; 0) \;,
\nonumber\\
\bar L_{3}(a) &\equiv& \bar C_{1} + \bar C_{3} =  \bar F(0; 1; -1;
0) \;, \,
\nonumber\\
\bar L_{4}(a) &\equiv&  2 \bar A_{1} + \bar A_{2} = \bar F(1; 1;
0; 0) \;, \,
\nonumber\\
\bar L_{5}(a) &\equiv& \bar D_{1} + \bar D_{3} = (1/4) \bar F(-1;
10; -17; 8) \;,
\nonumber\\
\bar L_{6}(a) &\equiv& (1/3) \bar D_{3} + \bar D_{6} =  (1/4) \bar
F(1; -12; 25; 28) \;,
\nonumber\\
\bar L_{7}(a) &\equiv& 3 \bar C_{1} - \bar A_{1} = (1/4) \bar
F(-1; -2; 3; 0) \;,
\nonumber\\
\bar L_{8}(a) &\equiv& 6 \bar C_{3} - \bar A_{2} = (1/2) \bar
F(-1; 12; -15; 0) \;,
\label{M25}
\end{eqnarray}
and $ L_{k}^{(p)}(\omega) = (6 / \pi \omega^{p+1})
\int_{0}^{\omega} y^{p} \bar L_{k}(y^{2}) \,d y .$

\section{The model of the background turbulent convection}

A simple approximate model for the three-dimensional isotropic
Navier-Stokes turbulence is described by a two-point correlation
function of the velocity field $ f_{ij}(t,{\bf x},{\bf y}) =
\langle u_i(t,{\bf x}) u_j(t,{\bf y}) \rangle $ with the
Kolmogorov spectrum $ W(k) \propto k^{-q} $ and $ q = 5/3 .$ The
turbulent convection is determined not only by the turbulent
velocity field $ {\bf u}(t,{\bf x}) $ but the fluctuations of the
entropy $ s(t,{\bf x}) .$ This implies that for the description of
the turbulent convection one needs additional correlation
functions, e.g., the turbulent flux of entropy $ {\bf
\Phi}_{i}(t,{\bf x},{\bf y}) = \langle s(t,{\bf x}) u_i(t,{\bf y})
\rangle $ and the second moment of the entropy fluctuations $
\Theta(t,{\bf x},{\bf y}) = \langle s(t,{\bf x}) s(t,{\bf y})
\rangle .$ Note also that the turbulent convection is anisotropic.

Now we derive Eqs.~(\ref{CB8}) and (\ref{CB9}) for the correlation
functions $ f_{ij} $ and $ {\bf \Phi}_{i} .$ To this end, the
velocity $ {\bf u}_{\perp} $ is written as a sum of the vortical
and the potential components, {\em i.e.,} $ {\bf u}_{\perp} =
\bec{\nabla} {\bf \times} (\tilde C {\bf e}) +
\bec{\nabla}_{\perp} \tilde \phi ,$ where $ w \equiv (\bec{\nabla}
{\bf \times} {\bf u})_z = - \Delta_{\perp} \tilde C ,$ $ \,
\Delta_{\perp} \tilde \phi = \Lambda u_{z} - \partial u_{z} /
\partial z ,$ $ \, \bec{\nabla}_{\perp} = \bec{\nabla} - {\bf e}
({\bf e} \cdot \bec{\nabla}) .$ Thus, in $ {\bf k} $-space the
velocity $ {\bf u} $ is given by
\begin{eqnarray}
u_i({\bf k}) = k_{\perp}^{-2} [ k^2 e_{m} P_{im}({\bf k})
u_{z}({\bf k}) - i ({\bf e} {\bf \times} {\bf k})_{i} w({\bf k})]
\;, \label{C80}
\end{eqnarray}
where we neglected terms $ \sim O(\Lambda) .$ Multiplying
Eq.~(\ref{C80}) for $ u_i({\bf k}_1) $ by $ u_j({\bf k}_2) $ and
averaging over turbulent velocity field we obtain
\begin{eqnarray}
f_{ij}^{(0)}({\bf k}) &=& k_{\perp}^{-4} [k^4 f^{(0)}({\bf k})
e_{mn} P_{im}({\bf k}) P_{jn}({\bf k}) \nonumber
\\
&&  + ({\bf e} {\bf \times} {\bf k})_{i} ({\bf e} {\bf \times}
{\bf k})_{j} G^{(0)}({\bf k})] \;, \label{C100}
\end{eqnarray}
where we assumed the turbulent velocity field in the background
turbulent convection is non-helical. Now we use an identity
\begin{eqnarray}
(k / k_{\perp})^{2} e_{mn} P_{im}({\bf k}) P_{jn}({\bf k}) =
e_{ij} + k_{ij}^{\perp} - k_{ij}
\nonumber\\
 = P_{ij}({\bf k}) - P_{ij}^{\perp}({\bf k}_{\perp}) \;,
 \label{C101}
\end{eqnarray}
which can be derived from
\begin{eqnarray*}
k_z (k_z e_{ij} + e_{i} k_{j}^{\perp} + e_{j} k_{i}^{\perp}) =
k_{ij} k^2 - k_{ij}^{\perp} k_{\perp}^{2} \; .
\end{eqnarray*}
Here we also used the identity $ ({\bf k}_{\perp} {\bf \times}
{\bf e})_{i} ({\bf k}_{\perp} {\bf \times} {\bf e})_{j} =
k_{\perp}^{2} P_{ij}^{(\perp)}(k_{\perp}) .$ Substituting
Eq.~(\ref{C101}) into Eq.~(\ref{C100}) we obtain
\begin{eqnarray}
f_{ij}^{(0)}({\bf k}) &&= (k / k_{\perp})^{2} \{f^{(0)}({\bf k})
P_{ij}({\bf k})
\nonumber \\
&&  + [G^{(0)}({\bf k}) / k^2 - f^{(0)}({\bf k})]
P_{ij}^{\perp}({\bf k}_{\perp}) \} \;, \label{C102}
\end{eqnarray}
Thus two independent functions determine the correlation function
of the turbulent velocity field. In isotropic three-dimensional
turbulent flow $ G^{(0)}({\bf k}) / k^2 = f^{(0)}({\bf k}) $ and
the correlation function reads
\begin{eqnarray}
f_{ij}^{(0)}({\bf k}) = f_{\ast} W(k) P_{ij}({\bf k}) / 8 \pi k^2
\; . \label{C103}
\end{eqnarray}
In isotropic two-dimensional turbulent flow $ G^{(0)}({\bf k}) /
k^2 \gg f_{\ast} f^{(0)}({\bf k}) $ and the correlation function
is given by
\begin{eqnarray}
f_{ij}^{(0)}({\bf k}) = G^{(0)}({\bf k}) P_{ij}^{\perp}({\bf
k}_{\perp}) / 8 \pi k^2 k_{\perp}^{2} \; . \label{C104}
\end{eqnarray}
A simplest generalization of these correlation functions is an
assumption that $ G^{(0)}({\bf k}) / [k^2 f^{(0)}({\bf k})] - 1 =
\varepsilon = const $ and thus the correlation function $
f_{ij}^{(0)}({\bf k}) $ is given by Eq.~(\ref{CB8}). This
correlation function can be considered as a combination of
Eqs.~(\ref{C103}) and (\ref{C104}) for three-dimensional and
two-dimensional turbulence. When $ \varepsilon $ depends on the
wave vector ${\bf k} ,$ the correlation function $
f_{ij}^{(0)}({\bf k}) $ is determined by two spectrum functions.

Now we derive Eq.~(\ref{CB9}) for the turbulent flux of entropy.
Multiplying Eq.~(\ref{C80}) written for $ u_i({\bf k}_2) $ by $
s({\bf k}_1) $ and averaging over turbulent velocity field we
obtain Eq.~(\ref{CB9}). Multiplying Eq.~(\ref{CB9}) by $ i ({\bf
k}_{\perp} {\bf \times} {\bf e})_{i} $ we get
\begin{eqnarray}
F^{(0)}({\bf k}) = i ({\bf k}_{\perp} {\bf \times} {\bf e}) \cdot
{\bf \Phi}_{\perp}^{(0)}({\bf k})  \; . \label{C105}
\end{eqnarray}
Now we assume that $ {\bf \Phi}_{\perp}^{(0)}({\bf k}) \propto
{\bf \Phi}^{\ast}_{\perp} f^{(0)}({\bf k}) / f_{\ast} .$ The
integration in $ {\bf k} $-space in Eq.~(\ref{C105}) yields the
numerical factor in Eq.~(\ref{CB11}). Note that for simplicity we
assumed that the correlation functions $ F^{(0)}({\bf k}) $ and $
f^{(0)}({\bf k}) $ have the same spectrum. If these functions have
different spectra, it results only in a different magnitude of a
numerical coefficient in Eq.~(\ref{CB11}).

Now let us discuss the physical meaning of the parameter   $
\sigma .$ To this end we will derive the equation for the
two-point correlation function $ \Phi_{z}^{(0)}({\bf r}) = \langle
s({\bf x}) \, {\bf u}({\bf x}+{\bf r}) \rangle $ of the turbulent
flux of entropy for the background turbulent convection (which
corresponds to Eq.~(\ref{CB10}) written in ${\bf k}$-space). To
this end we rewrite Eq.~(\ref{CB10}) in the following form:
\begin{eqnarray}
\Phi_{z}^{(0)}({\bf k}) &=& \Phi^{\ast}_z [k^{2} + \Gamma ({\bf e}
\cdot {\bf k})^{2}] \tilde \Phi_{w}(k) \;,
\label{CD1}\\
\tilde \Phi_{w}(k) &=& - (3 - \sigma) W(k) / 8 \pi k^{4} \;,
\label{CD2}
\end{eqnarray}
where $ \Phi^{\ast}_z = {\bf \Phi}^{\ast} \cdot {\bf e} ,$ $ \,
\Gamma = 3 (\sigma - 1) / (3 - \sigma) .$ The Fourier
transformation of Eq.~(\ref{CD1}) yields
\begin{eqnarray}
\Phi_{z}^{(0)}({\bf r}) &=& \Phi^{\ast}_z [\Delta + \Gamma ({\bf
e} \cdot \bec{\nabla})^{2}] \Phi_{w}(r) \;, \label{CD3}
\end{eqnarray}
where $ \Phi_{w}(r) $ is the Fourier transformation of the
function $ \tilde \Phi_{w}(k) .$ Now we use the identity
\begin{eqnarray}
\nabla_{i} \nabla_{j} \Phi_{w}(r) = \tilde \psi(r) \, \delta_{ij}
+ r \tilde \psi'(r) \, r_{ij} \;, \label{CD4}
\end{eqnarray}
where $ \tilde \psi(r) = r^{-1} \Phi'_{w}(r) $ and $
\tilde\psi'(r) = d \tilde\psi / dr .$ Equations~(\ref{CD3}) and
(\ref{CD4}) yield the two-point correlation function $
\Phi_{z}^{(0)}({\bf r}) :$
\begin{eqnarray}
\Phi_{z}^{(0)}({\bf r}) &=& \Phi^{\ast}_z \biggl(\tilde\psi(r) + r
\tilde\psi'(r) {1 + \Gamma \cos^{2} \tilde \theta \over 3 +
\Gamma} \biggr)  \;, \label{CD5}
\end{eqnarray}
where $ \tilde \theta $ is the angle between $ {\bf e} $ and $
{\bf r} .$ The function $ \tilde\psi(r) $ has the following
properties: $ \tilde\psi(r=0) = 1 $ and $ (r \tilde\psi')_{r=0} =
0 ,$ e.g., the function $ \tilde\psi(r) = 1 - (r / l_{0})^{q-1} $
satisfies the above properties, where $ 1 < q < 3 .$ Thus, the
two-point correlation function $ \Phi_{z}^{(0)}({\bf r}) $ of the
flux of entropy for the background turbulent convection is given
by
\begin{eqnarray*}
\Phi_{z}^{(0)}({\bf r}) = \Phi^{\ast}_z \biggl[1 &-& \biggl({(q -
1) (1 + \Gamma \cos^{2} \tilde \theta) \over 3 + \Gamma}
\\
&& + 1 \biggr) \biggl({r \over l_{0}} \biggr)^{q-1} \biggr] \;,
\end{eqnarray*}
where $ 1 < q < 3 .$ The simple analysis shows that $ - 3 / (q-1)
< \sigma < 3 ,$ where we took into account that $ \partial
\Phi_{z}^{(0)}({\bf r}) / \partial r < 0 $ for all angles $ \tilde
\theta .$ The parameter $ \sigma $ can be presented in the form $
\sigma = [1 + \tilde \xi (q + 1) / (q - 1)] / (1 + \tilde \xi / 3)
, $ where $ \tilde \xi  = (l_{\perp} / l_{z})^{q-1} - 1 ,$ $ \,
l_{\perp} $ and $ l_{z} $ are the horizontal $ (\tilde \theta =
\pi / 2) $ and vertical $ (\tilde \theta = 0) $ scales in which
the correlation function $ \Phi_{z}^{(0)}({\bf r}) $ tends to
zero. The parameter $ \tilde\xi $ describes the degree of thermal
anisotropy. In particular, when $ l_{\perp} = l_{z} $ the
parameter $ \tilde\xi = 0 $ and $ \sigma = 1 .$ For $ l_{\perp}
\ll l_{z} $ the parameter $ \tilde\xi = - 1 $ and $ \sigma = - 3 /
(q-1) .$ The maximum value $ \tilde \xi_{\rm max} $ of the
parameter $ \tilde \xi $ is given by $ \tilde \xi_{\rm max} = q -
1 $ for $ \sigma = 3 .$ Thus, for $ \sigma < 1 $ the thermal
structures have the form of column or thermal jets $ (l_{\perp} <
l_{z}) ,$ and $ \sigma > 1 $ there exist the `'pancake'' thermal
structures $ (l_{\perp} > l_{z}) $ in the background turbulent
convection.

\end{document}